\documentclass[%
 pre,
 amsmath,amssymb,
 reprint,%
]{revtex4-1}

\usepackage{graphicx}
\usepackage{dcolumn}
\usepackage{bm}

\usepackage[utf8]{inputenc}
\usepackage[T1]{fontenc}
\usepackage{mathptmx}
\usepackage{color}

\usepackage{algpseudocode}
\usepackage{algorithm}
\usepackage{varwidth}
\usepackage{hyperref}

\newcommand{\corr}{}

\newcommand{\cb}{}

\begin{document}

\preprint{AIP/123-QED}

\title{A numerical-continuation-enhanced flexible boundary condition scheme \\ applied to Mode I and Mode III fracture}

\author{Maciej Buze}
\email{BuzeM@cardiff.ac.uk}
\affiliation{ 
School of Mathematics, Cardiff University, Senghennydd Road, Cardiff, CF24 4AG, United Kingdom
}%

\author{James R. Kermode}
\email{J.R.Kermode@warwick.ac.uk}
\affiliation{%
Warwick Centre for Predictive Modelling, School of Engineering, University of Warwick, Coventry CV4 7AL, United Kingdom
}

\raggedbottom

\date{February 7, 2021}

\begin{abstract}
\noindent Motivated by the inadequacy of conducting atomistic simulations of crack propagation using static boundary conditions that do not reflect the movement of the crack tip, we extend Sinclair's flexible boundary condition algorithm [{\em Philos. Mag.} {\bf 31}, 647–671 (1975)] {\cb and propose a numerical-continuation-enhanced flexible boundary (NCFlex) scheme,} enabling full solution paths for cracks to be computed with pseudo-arclength continuation, and present a method for incorporating more detailed far-field information into the model for next to no additional computational cost. The new algorithms are ideally suited to study details of lattice trapping barriers to brittle fracture and can be incorporated into density functional theory and multiscale quantum/classical QM/MM calculations. We demonstrate our approach for Mode III fracture with a 2D toy model and {\cb employ it to conduct a 3D study of} Mode I fracture of silicon using realistic interatomic potentials, highlighting the superiority of the new approach over employing a corresponding static boundary condition. In particular, the inclusion of numerical continuation enables converged results to be obtained with realistic model systems containing a few thousand atoms, with very few iterations required to compute each new solution.
We also introduce a  method to estimate the lattice trapping range of admissible stress intensity factors $K_- < K < K_+$ very cheaply and demonstrate its utility on both the toy and realistic model systems.
\end{abstract}

\maketitle

\section{\label{sec:level1}Introduction}
{\cb The fundamental details of crack propagation invariably depend on atomistic effects, since a crack advances by the breaking of individual chemical bonds at its tip.}
Atomistic modelling of brittle fracture in crystals goes back to the pioneering work carried out by Sinclair and coworkers in the 1970s~\cite{Sinclair1972,Sinclair1972b,Sinclair1975}; for a recent review of contributions made to our understanding of fracture from atomistic simulations see Refs.~\onlinecite{Bitzek2015} and \onlinecite{Marder2016}.
The principal distinction from continuum models is the discreteness of the atomic lattice, which leads to the concept of lattice trapping, first identified by Thomson in 1971~\cite{Thomson1971}. A consequence of lattice trapping is that cracks remain stable over a range of stress intensity factors $K_- < K < K_+$.
Lattice trapping can lead to anisotropy in propagation directions~\cite{Perez2000}, and the associated energy barriers imply that cleavage does not necessary produce smooth fracture surfaces at low energies~\cite{Gumbsch2000}. The phenomenon has a dynamical analogue, the velocity gap, which is a forbidden band of crack velocities at low temperatures~\cite{Slepyan1981}. The velocity gap vanishes at larger temperatures; thermal activation over lattice trapping barriers has been proposed as an explanation for observations of low speed crack propagation on the $(110)$ cleavage plane in silicon~\cite{Kermode2015}.  
%
{\cb Very recently, bond-by-bond thermally activated crack growth has been directly observed in ReS$_2$ through \emph{in situ} atomic-resolution TEM experiments~\cite{Huang2020}, confirming the importance of understanding detailed atomistic mechanisms to control crack propagation.}

Detailed investigation of these phenomena are currently extremely challenging for two interconnected reasons. Firstly, realistic interatomic potentials capable of describing the very high strains near crack tips are very hard to construct~\cite{Holland1998_org}. At the same time, the requirement for large model systems and the strong coupling between lengthscales associated with fracture make the applicaton of quantum mechanical techniques such as density functional theory (DFT) extremely challenging, despite the considerable success such techniques have enjoyed elsewhere in materials science~\cite{Kermode2008}. Even when accurate atomistic models are available, determining the relevant stable crack tip configurations and the energy pathways that link them is extremely challenging because of the high dimensionality of the atomistic configuration space~\cite{Kermode2015}. The picture is further complicated if the modelled crack propagates, causing an effective shift of origin of the entire strain field, which is often not reflected in the supplied boundary condition. {\cb Challenges present in the (quasi-) static modelling of fracture translate directly to more complex simulations of fracture-related phenomenona, as pointed out in a recent study establishing the ill-suitability of a popular empirical potential for molecular dynamics  studies  of fracture phenomena in FeP  metallic  glasses \cite{He2019}.}

{\cb Many of the stated issues can be either resolved or significantly alleviated through supplementing the currently employed methods with ideas originating from numerical continuation and bifurcation theory, at present almost entirely absent from atomistic studies of material behaviour. Numerical continuation techniques concern efficient ways of computing solutions of a system of nonlinear equations by exploiting small variations in a parameter present in the system. Results from bifurcation theory ensure that this procedure can account for the changing stability of computed solutions -- see Ref.~\onlinecite{allgower2003introduction} for an accessible overview of the topic. These techniques allow the fracture predictions of a range of candidate interatomic potentials to be efficiently screened, helping to address issues such as that identified in \cite{He2019}; this approach will be demonstrated here using two potentials for silicon.

In our study we identify the stress intensity factor $K$ as such a parameter and propose a novel numerical-continuation-enhanced flexible boundary (NCFlex) scheme, which uses \emph{pseudo-arclength continuation} to trace continuous paths of equilibrium configurations, while employing flexible boundary conditions. As a result unstable solutions known as saddle points can be found, which constitute the energy barriers to crack propagation. A simpler version of this idea has been previously applied to study crack propagation in Refs.~\onlinecite{Li2013,Li2014}, albeit without explicitly identifying $K$ as the key parameter and without a recourse to the flexible boundary conditions.} Relatedly, a mathematically rigorous numerical analysis of domain size effects for a static boundary condition scheme coupled with numerical continuation techniques has been conducted in Ref.~\onlinecite{2019-antiplanecrack}. It provides a basic framework in which proving convergence rates to the infinite limit is possible and is the principal motivation for the current work. 

{\cb As mentioned,} here we return to the flexible boundary condition (FBC) approach introduced by Sinclair~\cite{Sinclair1975}. In this class of approaches, a localised atomic core region is coupled to a linear elastic far field{\cb , with only a few scalar parameters defining the far-field behaviour}. The method has been developed and applied extensively to model  dislocations~\cite{Sinclair1978,Yasi2012,Tan2016} but applications to fracture have received comparatively little attention{\corr , with the notable exception of work in Refs.~\onlinecite{Li2009,Li2012}, in which an efficient implementation of the FBC method in the context of both dislocations and cracks has been proposed.} We augment the FBC method and address the challenge of identifying and analysing stable and unstable crack tip configurations by combining it with numerical continuation techniques.
We demonstrate our ideas firstly for Mode III fracture in a toy model of a 2D crystal, considered a useful stepping stone for our theory, as it has readily calculable exact Hessians and permits a mathematically rigorous analysis. This is then followed by {\cb a 3D study of} the more realistic and much-studied example of Mode I fracture of silicon on the $(111)$ cleavage plane, using bond order potentials that have been modified to extend the interaction range and introduce screening to provide a qualitatively correct description of bond-breaking processes~\cite{Pastewka2013}.
\section{Methodology}
\subsection{Discrete Kinematics}
For the purpose of describing the method, we first consider a simplified system consisting of a two-dimensional infinite crystal of atoms forming a triangular lattice and interacting via a known interatomic potential with a finite interaction radius, with a crack forming along the horizontal axis and a crack tip located at $(\alpha,0) \in \mathbb{R}^2$. {\cb We stress, however, that the method is fully three-dimensional and the simplified setup is used as a backdrop to present the underlying ideas with clarity. The numerical example in Section \ref{sec:num_toy} follows this simplified setup, but the realistic study in Section \ref{sec:num_real} employs a fully three dimensional setup.}

The position of the $i$th atom is denoted by ${\bm{x}(i) = (x_1(i),x_2(i),x_3(i)) \in \mathbb{R}^3}$, which is always of the form
\begin{equation}\label{positions_form}
\bm{x}(i) = \bm{\hat{x}}(i) + {\cb\bm{Y}}(i),
\end{equation}
where $\bm{\hat{x}}(i)$ is the crystalline lattice position and ${\cb{\bm{Y}}(i) = (Y_1(i),Y_2(i),Y_3(i))}$ is the displacement from the crystalline lattice. The theory will be presented for two crack modes: pure Mode III {\cb in the out-of-plane displacement approximation} (${Y_1 = Y_2 = 0}$), and Mode I {\cb  in the in-plane displacement approximation} ($Y_3=0$). {\cb The realistic study in Section \ref{sec:num_real} does not involve such an approximation.} 

Atoms are assumed to interact according to an interatomic potential $\phi$, which, to avoid unnecessary technicalities is taken to be a pair potential with total energy of the form
\begin{equation}\label{phi}
E = \sum_{i\neq j} \phi(r_{ij}),\;\text{ where } r_{ij} = |{\cb \bm{x}}(i)-{\cb \bm{x}}(j)|.
\end{equation}
The restriction to pair potentials is not needed for the analysis, and will be lifted in the numerical examples considered in Section \ref{sec:num_real}{\cb , where a state-of-the-art many-body interatomic potential is used instead.}

\begin{figure}[!htb]
\includegraphics[width=\columnwidth]{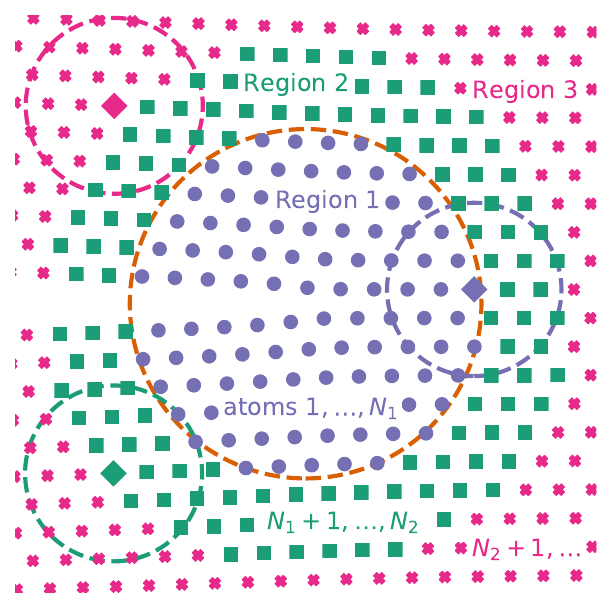}
\caption{\label{fig:regions} {\cb An example of {\corr the partitioning of a cracked infinite crystal.} Region 1 (purple dots) consists of atoms indexed $1,\dots, N_1$, with the shape for simplicity chosen to be a ball of radius $R$ ({\corr central} orange dotted line). Region 2 (green squares) consists of atoms indexed $N_1 + 1,\dots, N_2$ with its width defined by the interaction range. Region 3 (pink crosses) consists of atoms indexed $N_2 +1, \dots$ and it is the (in principle infinite) far field. A representative atom from each region is shown as an enlarged diamond and its interaction range is shown {\corr as a dotted line}. Note that every atom in Region 2 has at least one atom from Region 1 in its interaction range and the interaction range of every atom in Region 3 does not include any atoms from Region 1.} }
\end{figure}

Following the ideas of Sinclair \cite{Sinclair1975}, the system is divided into three regions{\cb , as shown in Figure \ref{fig:regions}.} Region {\cb 1}, also known as {\it the defect core }is a finite collection of atoms, labelled from $i=1$ to $i=N_1$, in the vicinity of the crack tip. Each atom in Region {\cb 1} is free to move, thus, in Mode I {\cb in the in-plane approximation}, there are $2N_1$ degrees of freedom associated with Region {\cb 1} and, in Mode III {\cb in the out-of-plane approximation}, there are $N_1$ degrees of freedom. {\cb The method can be easily adopted to a three dimensional setting, as will be done in Section \ref{sec:num_real}.}

Atoms in Region {\cb 2}, known as {\it the interface}, all contain at least one atom from Region {\cb 1} in its interaction range and are labelled from $i=N_1 +1$ to $i=N_2$. Region {\cb 3} is {\it the far field} {\cb and, by design, there is no interaction between atoms in Region 3 and 1. In models with long-range interactions such as DFT,  Region 2 should be, in theory, infinitely long. In practice, however, where there are long-range forces, Region 2 should be made sufficiently large that these forces can be truncated, with some control over the resulting errors. The output quantities $K_c, K_-, K_+$ would then effectively have to be converged with respect to the width of the interface.}

The total energy is divided into two parts 
\begin{equation}\label{energy_decomposition}
E = E^{(1)}(\{\bm{x}(i)\}_{i=1}^{N_2}) + E^{(2)}(\{\bm{x}(i)\}_{i > N_1}),
\end{equation}
{\cb where $E^{(1)}$ contains the energy of atoms in Region 1 and the energy associated with the interaction of atoms in Region 1 with atoms in Region 2 (hence the dependence for ${i = N_1 + 1,\dots,N_2}$). The second term, $E^{(2)}$, contains the energy of associated with atoms outside Region 1 -- this includes interactions between two atoms in Region 2, thus explaining the dependence for $i > N_1$. As discussed in Ref.~\onlinecite{Sinclair1975} and more recently in more general context in Ref.~\onlinecite{2018-antiplanecrack}, even if the energy is redefined so that the perfect lattice is a zero-energy configuration, the far-field part represented by $E^{(2)}$ nonetheless remains unbounded for a body containing a crack opening}. In practical applications the quantity of interest is thus the energy difference between a suitably chosen initial configuration $\{\bm{x_0}(i)\}$ and a relaxed configuration $\{\bm{x}(i)\}$, which is denoted by 
\begin{equation}\label{energy-diff-new}
E - E_0 = E(\{\bm{x}(i)\}) - E(\{\bm{x_0}(i)\}).
\end{equation}

There are different ways of specifying the behaviour of atoms in Region {\cb 2} and {\cb 3}. In what follows we first review two approaches, namely a simple static boundary condition specified by continuum linearised elasticity and a simplified version of the flexible boundary scheme due to Sinclair~\cite{Sinclair1975}.

Subsequently, we show how the flexible boundary scheme leads to a simple equation to check for admissible values of the stress intensity factor for which equilibria exist, which motivates defining an alternative version of the flexible boundary scheme with improved accuracy. 

This is then followed by a discussion about applying numerical continuation techniques to both formulations and the resulting bifurcation diagrams capturing crack propagation, energy barriers and the phenomenon of lattice trapping~\cite{Thomson1971}.

\subsection{Static boundary scheme}
Prescribing a simple static far-field boundary condition consists of constraining atoms in both Region~{\cb 2} and Region~{\cb 3} to be displaced according to continuum linear elasticity (CLE) equations arising from the mode of crack considered, the Cauchy-Born relation and the interatomic potential employed. Crucially, these equations are derived for the crack tip fixed at the origin, i.e. with $\alpha = 0$ and remain unchanged even if a relaxation of the core region yields a different crack tip position. 

In both idealised modes of crack considered, it can be shown that the atomistic model posed on a triangular lattice gives rise to an isotropic CLE equation \cite{Ostoja2006}. In particular, with the crack tip assumed to coincide with the origin of the coordinate system and polar coordinates ${\bm{\hat{x}}(i) = r_i(\cos\theta_i,{\cb\sin}\theta_i)}$ employed, the anti-plane CLE displacement for Mode III crack is given by 
\begin{equation}\label{CLE3}
{\cb\bm{U}}^{III}_{\rm CLE}(i) =  \sqrt{r_i}\left(0,0,\sin \left(\theta_i / 2\right)\right), 
\end{equation}
whereas the isotropic in-plane CLE displacement for Mode I crack is given by
\begin{align}\label{CLE1}
{\cb\bm{U}}^{I}_{\rm CLE}(i) =  \sqrt{r_i}\Big(&3\cos(\theta_i/2) - \cos(3\theta_i/2),\\ &5\sin(\theta_i/2) - \sin(3\theta_i/2),\,0\Big)\nonumber.
\end{align}
{\cb In the 3D study conducted in Section~\ref{sec:num_real}, the in-plane anisotropic Mode I CLE solution will be used, however the model considered will allow a full 3D relaxation of Region 1.} The superscript in ${\cb\bm{U}}_{\rm CLE}$ distinguishing different crack modes is dropped whenever a distinction is not needed. The displacement fields we consider are in fact of the form  $\{K\,{\cb\bm{U}}_{\rm CLE}(i)\}$, where $K \in \mathbb{R}$ is the stress intensity factor. 

A suitable way of encoding this far-field behaviour is to consider configurations $\{\bm{x}(i)\}$ in the form
\begin{equation}\label{ucle+u}
\bm{x}(i) = \bm{\hat{x}}(i) + K{\cb\bm{U}}_{\rm CLE}(i) + {\cb\bm{U}}(i),
\end{equation}
where, as in \eqref{positions_form}, $\bm{\hat{x}}(i)$ is the crystalline lattice position and ${\cb\bm{U}(i)}$ the atomistic correction of the $i$th atom, accounting for the fact that atoms within Region~{\cb 1} are free to relax under the interatomic potential. This correction is constrained to satisfy ${\cb\bm{U}}(i) = \bm{0}$ for $i > N_1$, which ensures that atoms outside the core remain fixed at the CLE displacement field. 

In this framework the initial configuration $\{\bm{x}_0(i)\}$ against which the energy difference is computed corresponds to setting ${\cb\bm{U}}(i) = 0$ for all $i${\cb , with $K$ the same for both $\bm{x}$ and $\bm{x_0}$.} As a result, {\cb given the definition of $E^{(2)}$ in \eqref{energy_decomposition},} it trivially holds that 
\[
E{\cb^{(2)}}(\{\bm{x}(i)\}_{i > N_1}) = E{\cb^{(2)}}(\{\bm{x}_0(i)\}_{i > N_1}),
\]
and hence
\[
E-E_0 = E(\{\bm{x}(i)\}_{i=1}^{N_2}) - E(\{\bm{x_0}(i)\}_{i=1}^{N_2}),
\]
which is a finite quantity. 

If one defines a function of $(\{{\cb\bm{U}}(i)\}, K)$ given by 
\begin{equation}\label{F0}
\bm{F_0} = {\cb \bm{F_0}((\{{\cb\bm{U}}(i)\}, K))} = (\{\bm{f}(i)\}_{i=1}^{N_1}),
\end{equation}
where $\bm{f}(i) = -\frac{\partial E}{\partial \bm{x}(i)}$ is the force acting on the $i$th atom, then an equilibrium configuration can be found by solving $\bm{F_0} = \bm{0}$. {\cb The usual numerical procedure is to prescribe some reasonable $K$ and check whether there exists a solution to $\bm{F_0} = \bm{0}$ for that $K$.} For a fixed $K$, in Mode III, {\cb this corresponds to solving} a system of $N_1$ equations for $N_1$ variables, and, in Mode I there are  $2N_1$ equations for $2N_1$ variables. In a fully 3D case, to be considered in numerical tests in Section \ref{sec:num_real}, the system considered consists of $3N_1$ equations for $3N_1$ variables.

The remaining difficulty is the interplay between the choice of $K$ and the crack tip position. This will be addressed in Section \ref{sec:cont} with the help of numerical continuation, in particular highlighting fundamental limitations of the static boundary scheme. 

\subsection{Flexible boundary scheme}
\subsubsection{Standard formulation}
The central idea of the flexible boundary scheme described by Sinclair~\cite{Sinclair1975} is to allow the crack tip position $(\alpha,0)$ to vary. The displacements can be shifted to account for the current crack tip position by redefining the polar coordinates used in \eqref{CLE3} and \eqref{CLE1} so that
\[
r_i(\cos\theta_i,\sin\theta_i) = \bm{\hat{x}}(i) - (\alpha,0).
\]
The configurations considered are, similarly to \eqref{ucle+u}, of the form
\[
\bm{x}(i) = \bm{\hat{x}}(i) + K\,\cb{\bm{U}}^{\alpha}_{\rm CLE}(i) + {\cb\bm{U}}(i),
\]
where the CLE displacement is now written as ${\cb\bm{U}}_{\rm CLE}^{\alpha}$ to emphasise the dependence on $\alpha$ through the shift of the polar coordinate system. {\cb Note that }the initial unrelaxed configuration $\{\bm{x_0}(i)\}$ {\cb from \eqref{energy-diff-new} corresponds to} setting $\alpha = 0$ and ${\cb\bm{U}}(i) = 0$ for all $i${\cb , whereas $K$ is again the same for both $\bm{x}$ and $\bm{x_0}$. }

The effect that varying $\alpha$ has on the system can be captured by considering the notion of a {\it generalised force}
\begin{equation}\label{gen_force}
f_{\alpha}^{\infty} = -\frac{\partial E}{\partial \alpha} = \sum_{i}-\frac{\partial E}{\partial \bm{x}(i)}\cdot \frac{\partial \bm{x}(i)}{\partial \alpha} = \sum_i \bm{f}(i) \cdot K\, {\cb \bm{V}}_{\alpha}(i),
\end{equation}
where ${\cb\bm{V}}_{\alpha} = -\partial_{1} {\cb\bm{U}}^{\alpha}_{\rm CLE}${\cb , noting that, morally, $\bm{U}_{\rm CLE}^{\alpha}$ is a function from $\mathbb{R}^2 \to \mathbb{R}^3$, so one can write ${\cb\bm{U}}_{\rm CLE}^{\alpha}(\bm{x})$, where ${\bm{x} = (x_1,x_2) \in \mathbb{R}^2}$, and $\partial_1$ simply refers to the derivative with respect to $x_1$, which is consistent with the fact that we vary the horizontal position of the crack tip.} As stated, {\cb \eqref{gen_force} }is an infinite sum, which can be shown to be convergent since ${\cb\bm{U}}_{\rm CLE}$ solves the CLE equation~\cite{EOS2016,2018-antiplanecrack}.  

Somewhat arbitrarily, Sinclair assumes that in Region {\cb 3} the crystal is fully 'linearly elastic'~\cite{Sinclair1975}, in the sense that the continuum CLE displacement is an equilibrium by itself, meaning that 
\begin{equation}\label{forcesff}
\bm{f}(i) = 0,\text{ for }i > N_2,
\end{equation}
for any choice of $K$ and $\alpha$, effectively truncating the infinite sum in \eqref{gen_force}. 

The Sinclair scheme can be formalised by defining a function of $(\{{\cb\bm{U}}(i)\},K,\alpha\}\cb{)}$ given by 
\begin{equation}\label{F1}
\bm{F_1} = {\cb\bm{F_1}(\{{\cb\bm{U}}(i)\},K,\alpha\}) =} ( (\{\bm{f}(i)\}_{i=1}^{N_1}, f_{\alpha}), 
\end{equation}
where 
\begin{equation}\label{falpha}
f_{\alpha} = \sum_{i=1}^{N_2} \bm{f}(i) \cdot K\,{\cb\bm{V}}_{\alpha}(i).
\end{equation}
An equilibrium configuration in this scheme is then obtained by solving $\bm{F_1} = \bm{0}$ {\cb where $K$ is as again a priori fixed at some reasonable value}.

Notably, the summation in \eqref{falpha} is effectively over $\{i\}_{i=N_1+1}^{N_2}$, since at an equilibrium $\bm{f}(i) = 0$ for $i \leq N_1$. {\cb The reasoning behind including the extra condition $f_\alpha = 0$ can be explained as follows. With atoms outside Region 1 following the CLE displacement (determined by $K$ and $\alpha$), it can never be true that $\bm{f}(i) = 0$ for $i > N_1$ -- optimising over $\alpha$ is hence the best we can hope to achieve outside the defect core.}

Obviously, \eqref{forcesff} is only true in an approximate sense, which leaves open to interpretation whether the truncation enforced through \eqref{forcesff} is the optimal choice.  

It is further worth noting that in the limit when $N_1 \to \infty$, the generalised force $f_{\alpha}$ in \eqref{falpha} is null at any equilibrium, hence the extra equation $f_\alpha = 0$ is effectively redundant, which strongly hints that the role of the flexible scheme lies in improving the convergence rate to the single infinite limit. A mathematically rigorous proof of this result will be a subject of further study. 

To compute the energy difference ${E - E_0}$ in the new scheme, we follow the procedure described in Ref.~\onlinecite[Appendix 1]{Sinclair1975}, with the far-field contribution to the energy ${E^{(2)}-E^{(2)}_0}$ from \eqref{energy_decomposition} approximated as
\begin{equation}\label{E-ff-approx}
E^{(2)}-E^{(2)}_0 =  -\frac{1}{2}\sum_{i=N_1+1}^{N_2}\left(\bm{f}^{(2)}(i) + \bm{f_0}^{(2)}(i)\right)\cdot\left(\bm{x}(i)-\bm{x_0}(i)\right),
\end{equation}
where $\bm{f}^{(2)}(i) = -\frac{\partial E^{(2)}}{\partial \bm{x}(i)}$. {\cb We in particular note that in this approximation the contributions from atoms in Region 3 ($i > N_2$) are disregarded completely, due to the truncation in \eqref{forcesff}, which holds for both $\bm{f}^{(2)}$ and $\bm{f_0}^{(2)}$, and that the difference $\bm{x}(i)-\bm{x_0}(i)$ for $i > N_1$ is only a function $\alpha$.}

Making sense of the arguably unsubstantiated far-field approximation in \eqref{forcesff} as well as addressing the question of convergence leads to several interesting realisations that will be addressed in the next section.

\subsubsection{Predicting admissible stress intensity factors}\label{sec:pred_K_CLE}
The strain fields associated with atomistic corrections $\{{\cb\bm{U}}(i)\}$ are known to decay more quickly away from the core than the strain fields associated to $\{{\cb\bm{U}}_{\rm CLE}(i)\}$ (proven rigorously in a simplified setup in Ref.~\onlinecite{2018-antiplanecrack}), meaning that their contributions are effectively negligible beyond a small region around the crack tip. It can thus be conjectured that a reasonable approximation to the flexible boundary scheme condition $f_{\alpha} = 0$, defined in \eqref{falpha}, is to allow the unrelaxed configuration $\{\bm{x_0}(i)\}$ to depend on $\alpha$ and look at the generalised force at the unrelaxed configuration, namely
\[
-\frac{\partial E_0}{\partial \alpha} = \sum_{i}-\frac{\partial E_0}{\partial \bm{x_0}(i)}\cdot \frac{\partial \bm{x_0}(i)}{\partial \alpha} = \sum_i \bm{f_0}(i) \cdot K\, {\cb\bm{V}}_{\alpha}(i).
\]
Employing the same truncation as in \eqref{forcesff}, we can postulate a condition $f^0_{\alpha} = 0$, where
\begin{equation}\label{f0_alpha}
{f}^0_{\alpha}{\cb (K,\alpha)} = \sum_{i=1}^{N_2}\bm{f_0}(i) \cdot K\,{\cb\bm{V}}_{\alpha}(i).
\end{equation}
With the unrelaxed configuration $\{\bm{x_0}(i)\}$, now determined solely by $K$ and $\alpha$, verifying whether ${f}^0_{\alpha} = 0$ holds is numerically very straightforward { \cb and comes at minuscule computational cost. Note, however, that with $f_{\alpha}^0(i)$ representing the forces acting on the $i$th atom at the unrelaxed configuration $\bm{x_0}$, which depends on $K$, $f^0_{\alpha}$ exhibits a nonlinear dependence on $K$.}

It will be shown in Section~\ref{num_pred_k} and Section~\ref{sec:num_real} that solving ${f}^0_{\alpha} = 0$ provides a good estimate for the admissible values of the stress intensity factor and that there in fact exists a continuous path of solutions with $K$ values nearly perfectly oscillating around a fixed interval of admissible values ${K_- < K < K_+}$. 

With the numerical tests indicating that the predicted interval is strongly dependent on the size of the computational domain, it seems plausible that changing the far-field truncation rule from \eqref{forcesff} can have a drastic effect on the computed solution path. This will be investigated in the next section.  
\subsubsection{Effect of changing far-field truncation rule}\label{sec:new_flex}
\begin{figure}[!htb]
\includegraphics[width=\columnwidth]{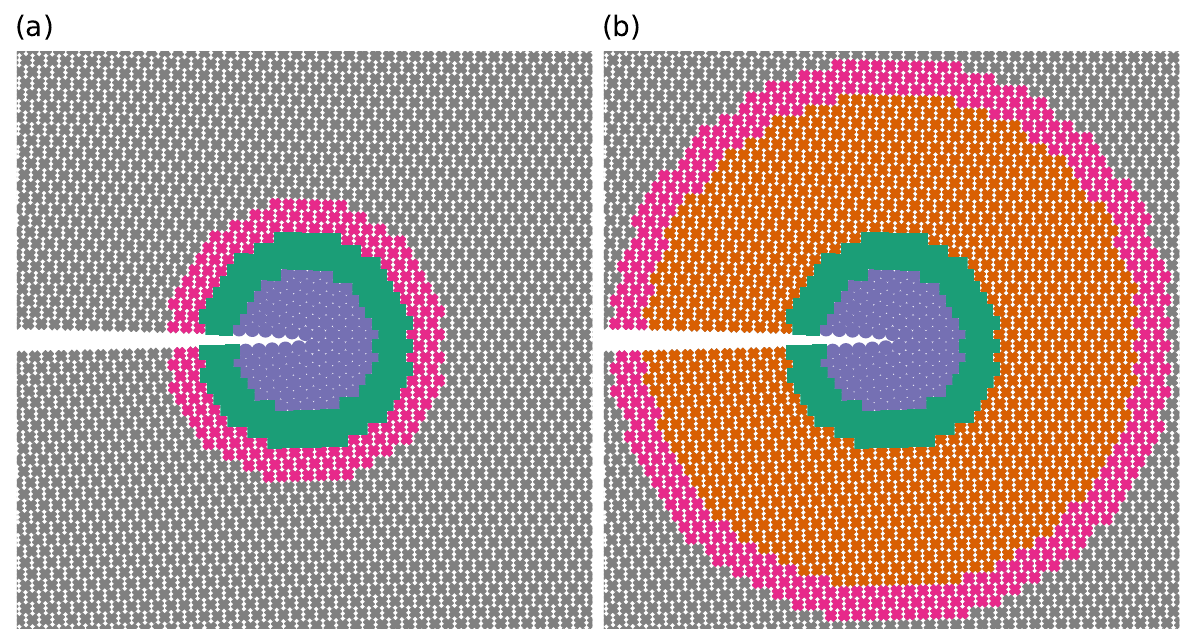}
\caption{\label{fig:regions_ff} {\cb The effective computational domain for {\corr (a)} the standard flexible scheme and {\corr (b)} the flexible scheme with extended far field, with colouring system matching the one in Figure \ref{fig:regions} {\corr and with Region~1 being the inner ball and the interface being the strip around it}. The standard truncation $\bm{f}(i) = 0$ for all $i > N_2$ from \eqref{forcesff} implies that only a narrow strip of Region 3 ({\corr the outermost strip in pink in (a)}) has to be simulated, so that $\bm{f}(i)$ can be computed for $i = N_1+1,\dots,N_2$, while the rest of the infinite far field ({\corr the outermost atoms} in grey) can be disregarded. If on the other hand the truncation is imposed for $i > N_3$ (as in \eqref{forcesff2}), where $N_3$ is much larger, then atoms $i=N_2+1,\dots,N_3$ have to also be simulated ({\corr the in-between thick strip} presented in orange {\corr in (b)}, but conceptually part of Region~3), as well as a further strip around then so that all relevant forces can be computed. Atoms in the orange and pink regions on the right still simply follow the CLE displacement determined by $K$ and $\alpha$, so the enlarged domain still contains the same number of degrees of freedom.}}
\end{figure}

The truncation in \eqref{forcesff} is equivalent to stating that the atomistic information associated with atoms in Region {\cb 3}, which conceptually is an infinite far field, is completely disregarded{\cb , except for tiny strip, as shown in Figure~\ref{fig:regions_ff}.} One can provide the flexible boundary scheme with more atomistic input from Region {\cb 3} by changing the truncation in \eqref{forcesff} to
\begin{equation}\label{forcesff2}
    \bm{f}(i) = 0,\;\text{ for }i > N_3,
\end{equation}
where $N_3 > N_2$ is much larger, e.g. $N_3 = 4 N_2$. {\cb Conceptually, this is equivalent to increasing the width of Region 2 (c.f. Figure \ref{fig:regions}), but in our phrasing $N_2$ is uniquely specified by $N_1$ and the interaction range of the potential, hence the need for the introduction of a much larger $N_3$ is simply dictated by the wording -- see Figure \ref{fig:regions_ff} for further insight.} 

As a result, a new condition is $\tilde{f}_{\alpha} = 0$, where
\[
\tilde{f}_{\alpha} = \sum_{i=1}^{N_3} \bm{f}(i)\cdot K\,{\cb\bm{V}}_{\alpha}(i).
\]
Thus, the new approach testing the effect of different truncation can be formalised by defining a function of $(\{{\cb\bm{U}}(i)\},K,\alpha)$ given by 
\begin{equation}\label{F2}
\bm{F_2} = {\cb \bm{F_2}(\{{\cb\bm{U}}(i)\},K,\alpha) =} (\{\bm{f}(i)\}_{i=1}^{N_1}, \tilde{f}_{\alpha}),
\end{equation}
with an equilibrium configuration obtained by solving $\bm{F_2} = 0$.

The difference between the new scheme $\bm{F_2} = 0$ and the standard $\bm{F_1} = 0$ from \eqref{F1} is most pronounced by observing that at an equilibrium in the new scheme, the previously defined truncated generalised force $f_{\alpha}$ from \eqref{falpha} satisfies 
\begin{equation}\label{falpha_new}
f_{\alpha} = \sum_{i=N_2+1}^{N_3}\bm{f}(i)\cdot K\,V_{\alpha}(i).
\end{equation}
The right-hand side admits input only from atoms in Region~{\cb 3}, whose displacements are determined solely by $\alpha$ and $K$, since by design, ${\cb\bm{U}}(i) = 0$ for  $i \geq N_1$, which highlights the general rationale behind this formulation: the far-field region within the computational domain is vastly enlarged, but only two degrees of freedom remain attached to it, meaning that in practice there is virtually no additional computational cost, apart from the ability to compute the right-hand side of \eqref{falpha_new}.

It will be shown through numerical tests presented in  Section~\ref{sec:num_new_flex} that the new scheme results in much improved accuracy for small sizes of the core region, implying that in practice the new scheme is numerically preferable, enabling increased accuracy to be achieved with decreased numerical cost.  

\subsection{Pseudo-arclength numerical continuation}\label{sec:cont}
The basic premise of numerical continuation applied to the problem at hand is as follows. Suppose we have identified some {\cb $K_n$} for which some equilibrium configuration {\cb $\bm{x}_n$} exists. Can we use this knowledge to quickly find another equilibrium for {\cb $K_n + \delta K$}, for some small $\delta K$? {\cb A similar approach has been previously applied in the continuum study of cracks deviating from straightness \cite{Rice1985}. In the present setting,} such an approach will work well if there exists a continuous path of solutions $K \mapsto \{{\cb\bm{U}}_K(i)
\}_{i=1}^{N_1}$ (and $K \mapsto \alpha_K$ in the case of the flexible boundary scheme). Such a path can be shown to exist, courtesy of Implicit Function Theorem~\cite{serge}, in the neighbourhood of {\cb $K_n$} if the associated Hessian operator is invertible at {\cb $K_n$}. 

A more sophisticated version, which is particularly useful for the problem at hand, is known as the pseudo-arclength continuation. It  postulates that the quantities involved all are smooth functions of an arclength parameter $s$. {\cb The arclength parametrisation of a curve, also known as the natural parametrisation, is a classical concept in mathematical analysis -- it ensures that the curve is traversed at a constant unit speed \cite{rudin1976principles}.} The question thus changes to: given some triplet {\cb $\left(K_{s_n},\{{\cb\bm{U}}_{s_n}(i)\},\alpha_{s_n}\right)$} (in the case of static boundary $\alpha \equiv 0$ throughout) which specifies an equilibrium configuration {\cb $\{\bm{x}_{s_n}(i)\}$}, can we find a new triplet for {\cb $s_{n+1} := s_n+\delta s$}, for some small $\delta s$, which gives us a new equilibrium {\cb $\{\bm{x}_{s_{n+1}}(i)\}$}? The key advantage of this approach is that it can handle index-1 saddle points, which makes it a useful tool for studying energy barriers and the phenomenon of lattice trapping.

Numerical continuation can be incorporated into the framework by including $K$ as a variable in the systems of equations $\bm{F_j} = 0$ for $j=0,1,2$ defined in \eqref{F0}, \eqref{F1}, \eqref{F2}. The inclusion of $K$ as a variable renders each system of equations  $\bm{F_j} = 0$ under-determined. {\cb While it might be tempting to optimise over $K$ by requiring that $\frac{\partial E}{\partial K} = 0$, as was the case for $\alpha$, we stress that the nature of $K$ is fundamentally different to $\alpha$ -- the value of $K$ specifies the dominant behaviour at infinity, scaling like $\sim r^{1/2}$ in the displacement and $\sim r^{-1/2}$ in the strain, where $r$ is the distance from the crack tip. Due to this dominant behaviour at infinity, it was shown in \cite{2019-antiplanecrack} that optimising over $K$ is not mathematically sound. On the other hand, $\alpha$ specifies the next order behaviour, with its contributions $\sim r^{-1/2}$ in the displacement and $\sim r^{-3/2}$ in the strain.  This is "small enough" at infinity to justify the constraint $\frac{\partial E}{\partial \alpha} = 0$.} 

\begin{figure}[!htb]
\includegraphics[trim=0 15 40 0, clip, width=0.85\columnwidth]{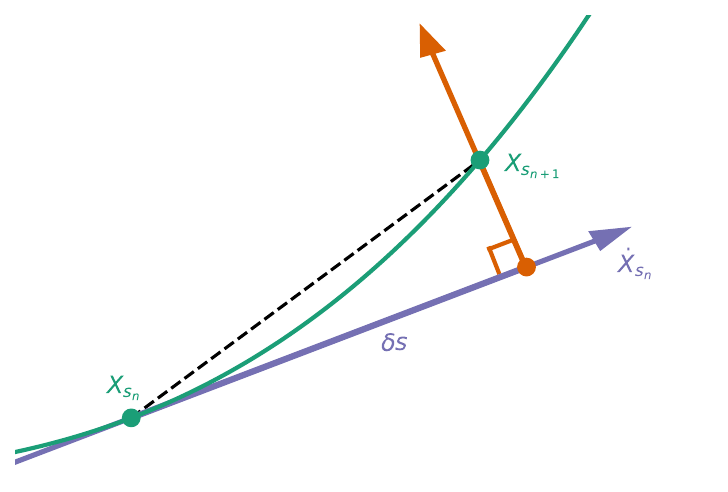}
\caption{\label{fig:schematic_plot} Suppose $X_{s_n} = (K_{s_n},\{{\cb\bm{U}}_{s_n}(i)\},\alpha_{s_n})$ belongs to the solution path, which is smooth (solid {\corr curved} green line). The schematic two-dimensional plot indicates that imposing an extra constraint in the form of {\cb $f^{(n)}_K = 0$} from \eqref{fk} is is equivalent to looking for solutions along the direction of the orange arrow {\corr pointing up}, which is perpendicular to $\dot{X}_{s_n}$, represented by a purple arrow {\corr pointing right}, with $\delta s$ determining how far along $\dot{X}_{s_n}$ we choose to travel. Choosing $\delta s$ small enough ensures that the algorithm can safely traverse the solution path even along folds. It is further clear from the plot that, for $\delta s$ sufficiently small, $X_{s_n} + \delta s \dot{X}_{s_n}$ (orange dot {\corr on the arrow pointing right}) is a very good initial guess for subsequent steps of the Newton iteration.}
\end{figure}

{\cb The principles of pseudo-arclength continuation, as introduced at the beginning of this section, dictate that, under the assumption of there existing a continuous path of solutions, {\cb given a solution triplet $(K_{s_n},\{\bm{U}_{s_n}(i)\},\alpha_{s_n})$,} one should instead impose the equation {\cb $f^{(n)}_K = 0$} to close the system, where 
\begin{align}\label{fk}
{\cb f^{(n)}_{\rm K}}= \sum_{i=1}^{N_1} &\left({\cb\bm{U}}_{s_{n+1}}(i) - {\cb\bm{U}}_{s_n}(i)\right) \cdot \dot{{\cb\bm{U}}}_{s_n}(i)\\ &+ (\alpha_{s_{n+1}} - \alpha_{s_n})\dot{\alpha}_{s_n} + (K_{s_{n+1}}-K_{s_n})\dot{K}_{s_n} - \delta s.\nonumber
\end{align}
{\cb The schematic plot presented in Figure \ref{fig:schematic_plot} explains this construction. }Here $\left(\dot{K}_{s_n},\{\dot{{\cb\bm{U}}}_{s_n}(i)\},\dot{\alpha}_{s_n}\right)$ refer to derivatives with respect to $s$ evaluated at $s_n$. Note that in the static boundary scheme we simply have $\alpha_{s_n} = 0$, and hence $\dot{\alpha}_{s_n} = 0$, for all $n$.}

If there indeed exist a continuous path of solutions, it can be shown~\cite{Beyn_numericalcontinuation} that for a step size $\delta s$ small enough, a simple Newton iteration will converge to a new solution of ${(\bm{F_j}, f_K) = \bm{0}}$. 

The remaining difficulty is to compute {\cb $\left(\dot{K}_{s_n},\{\dot{{\cb\bm{U}}}_{s_n}(i)\},\dot{\alpha}_{s_n}\right)$}. This can be achieved by first noting that with $s$ being an arclength parameter, by definition {\cb (unit speed)} it has to hold that 
\begin{equation}\label{s-arc}
{\cb     \sum_{i=1}^{N_1}\dot{{\cb\bm{U}}}_{s_n}(i) \cdot \dot{{\cb\bm{U}}}_{s_n}(i) + (\dot{K}_{s_n})^2 + (\dot{\alpha}_{s_n})^2 = 1. }
\end{equation}
This eliminates one degree of freedom. The remaining degrees of freedom can be eliminated by  differentiating both sides of $\bm{F}_j = 0$ with respect to arclength parameter $s$, which is possible under the assumption of there existing a smooth path of solutions. Details are presented in the \hyperref[app1]{Appendix}.

The resulting pseudo-arclength continuation algorithms associated with both schemes are presented as Algorithm \ref{alg1} and Algorithm \ref{alg2}.

\begin{algorithm}[H]
   \caption{Static boundary condition pseudo-arclength continuation}\label{alg1}
   \begin{algorithmic}[1]
    \State Given $\delta s$;
      \State given a \underline{stable} equilibrium configuration solving $(\bm{F_0},f_K) = \bm{0}$ determined by $(K_{s_1},\{{\cb\bm{U}}_{s_1}(i)\})$;
      \State compute $(\dot{K}_{s_1},\{\dot{{\cb\bm{U}}}_{s_1}(i)\})$ using \eqref{udot_kdot};
      \State compute a new stable equilibrium configuration $(K_{s_2},\{{\cb\bm{U}}_{s_2}(i)\})$ by solving $(\bm{F_0},f_K) = \bm{0}$ using Newton iteration with initial guess $(K_{s_1}+ \delta s \dot{K}_{s_1},\{{\cb\bm{U}}_{s_1}(i) + \delta s \dot{{\cb\bm{U}}}_{s_1}(i)\})$;
      
      \For{$n > 1$}
      \State given $(K_{s_{n}},\{{\cb\bm{U}}_{s_{n}}(i)\})$ and $(\dot{K}_{s_{n-1}},\{\dot{{\cb\bm{U}}}_{s_{n-1}}(i)\})$;
      \State compute $(\dot{K}_{s_n},\{\dot{{\cb\bm{U}}}_{s_n}(i)\})$ by solving linear system \eqref{dsF0_approx};
      
      \State \begin{varwidth}[t]{0.88\linewidth} compute a new equilibrium configuration $(K_{s_{n+1}},\{{\cb\bm{U}}_{s_{n+1}}(i)\})$ by solving $(\bm{F_0},f_K) = \bm{0}$ using Newton iteration with initial guess $(K_{s_n}+ \delta s \dot{K}_{s_n},\{{\cb\bm{U}}_{s_n}(i) + \delta s \dot{{\cb\bm{U}}}_{s_n}(i)\})$. \end{varwidth}
  \EndFor
\end{algorithmic}
\end{algorithm}

\begin{algorithm}[H]
   \caption{Flexible boundary condition pseudo-arclength continuation}\label{alg2}
   \begin{algorithmic}[1]
    \State Given $\delta s$;
      \State given a \underline{stable} equilibrium configuration solving $(\bm{F_1},f_K) = \bm{0}$ determined by $(K_{s_1},\{{\cb\bm{U}}_{s_1}(i)\},\alpha_{s_1})$;
      \State compute $(\dot{K}_{s_1},\{\dot{{\cb\bm{U}}}_{s_1}(i)\},\dot{\alpha}_{s_1})$ using \eqref{udot_kdot_alphadot};
      \State compute a new stable equilibrium configuration $(K_{s_2},\{{\cb\bm{U}}_{s_2}(i)\},\alpha_{s_2})$ by solving $(\bm{F_1},f_K) = \bm{0}$ using Newton iteration with initial guess $(K_{s_1}+ \delta s \dot{K}_{s_1},\{{\cb\bm{U}}_{s_1}(i) + \delta s \dot{{\cb\bm{U}}}_{s_1}(i)\},\alpha_{s_1} + \delta s \dot{\alpha}_{s_1})$;
      
      \For{$n > 1$}
      \State given $(K_{s_{n}},\{{\cb\bm{U}}_{s_{n}}(i)\},\alpha_{s_n})$ and $(\dot{K}_{s_{n-1}},\{\dot{{\cb\bm{U}}}_{s_{n-1}}(i)\},\dot{\alpha}_{s_{n-1}})$;
      \State compute $(\dot{K}_{s_n},\{\dot{{\cb\bm{U}}}_{s_n}(i)\},\dot{\alpha}_{s_n})$ by solving linear system \eqref{dsF1_approx};
      
      \State \begin{varwidth}[t]{0.88\linewidth}\sloppy compute a new equilibrium configuration $(K_{s_{n+1}},\{{\cb\bm{U}}_{s_{n+1}}(i)\},\alpha_{s_{n+1}})$ by solving ${(\bm{F_1},f_K) = \bm{0}}$ using Newton iteration with initial guess ${(K_{s_n}+ \delta s \dot{K}_{s_n},\{{\cb\bm{U}}_{s_n}(i) + \delta s \dot{{\cb\bm{U}}}_{s_n}(i)\},\alpha_{s_n} + \delta s \dot{\alpha}_{s_n})}$. \end{varwidth}
  \EndFor
\end{algorithmic}
\end{algorithm}

Bearing in mind that most realistic interatomic potentials only provide analytic forces but not Hessians, meaning that Algorithm \ref{alg2} cannot be readily used, as it requires a computation of the Hessian while differentiating $\bm{F_j}=0$ to get the tangent $(\dot{K}_{s_n},\dot{{\cb\bm{U}}}_{s_n},\dot{\alpha}_{s_n})$, we also propose a simple finite-difference based approximate scheme as a Hessian-free alternative.

The method consists of first computing two stable equilibrium configurations determined by $(K_{s_0},\{{\cb\bm{U}}_{s_0}(i)\},\alpha_{s_0})$ and $(K_{s_1},\{{\cb\bm{U}}_{s_1}(i)\},\alpha_{s_1})$, which crucially satisfy $K_{s_1} \approx K_{s_0}$ (and also $\alpha_{s_1} \approx \alpha_{s_0}$). In the first step the tangent $(\dot{K}_{s_1},\{\dot{{\cb\bm{U}}}_{s_1}(i)\},\dot{\alpha}_{s_1})$ can be approximated as 
\begin{subequations}\label{dot_approx1}
\begin{align}
    \dot{{\cb\bm{U}}}_{s_1}(i) &= \frac{1}{K_{s_1}-K_{s_0}}\left({\cb\bm{U}}_{s_1}(i) - {\cb\bm{U}}_{s_0}(i)\right),\\
    \dot{\alpha}_{s_1} &= \frac{1}{K_{s_1}-K_{s_0}}\left(\alpha_{s_1} - \alpha_{s_0}\right),\\
    \dot{K}_{s_1} &= 1,
\end{align}
\end{subequations}
with the last line a direct consequence of $K$ being the effective continuation parameter in the first step, since it is $K$ that is varied to obtain two stable equilibrium configurations.

With the tangents computed, one can now assemble the extended system and solve $(\bm{F_1},f_K) = \bm{0}$ to obtain an equilibrium determined by $(K_{s_2},\{{\cb\bm{U}}_{s_2}(i)\},\alpha_{s_2})$. The switch to the extended system entails that now the arclength $s$ is the continuation parameter and in particular $s_2 - s_1 = \delta s$, with $\delta s$ fixed throughout. As a result, subsequent tangent approximations are computed, for $n=2,\dots$, as 
\begin{subequations}\label{dot_approx2}
\begin{align}
    \dot{{\cb\bm{U}}}_{s_{n}}(i) &= \frac{1}{\delta s}\left({\cb\bm{U}}_{s_n}(i) - {\cb\bm{U}}_{s_{n-1}}(i)\right),\\
    \dot{\alpha}_{s_n} &= \frac{1}{\delta s}\left(\alpha_{s_n} - \alpha_{s_{n-1}}\right),\\
    \dot{K}_{s_n} &= \frac{1}{\delta s}\left(K_{s_n} - K_{s_{n-1}}\right).
\end{align}
\end{subequations}
The details of this approximate scheme are summarised as Algorithm \ref{alg3} below. In practical applications, to avoid possible numerical artefacts, the finite-difference approach could be substituted by the automatic differentiation approach~\cite{Neidinger2010}.

\begin{algorithm}[H]
   \caption{Hessian-free approximate flexible boundary condition pseudo-arclength continuation}\label{alg3}
   \begin{algorithmic}[1]
    \State Given $\delta s$;
      \State given two \underline{stable} equilibrium configurations solving $(\bm{F_1},f_K) = \bm{0}$, determined by $(K_{s_0},\{{\cb\bm{U}}_{s_0}(i)\},\alpha_{s_0})$  and $(K_{s_1},\{{\cb\bm{U}}_{s_1}(i)\},\alpha_{s_1})$, and satisfying $K_{s_1} \approx K_{s_0}$ and $\alpha_{s_1} \approx \alpha_{s_0}$;
      \State compute an approximate $(\dot{K}_{s_1},\{\dot{{\cb\bm{U}}}_{s_1}(i)\},\dot{\alpha}_{s_1})$ using \eqref{dot_approx1};
      \State compute a new stable equilibrium configuration $(K_{s_2},\{{\cb\bm{U}}_{s_2}(i)\},\alpha_{s_2})$ by solving $(\bm{F_1},f_K) = \bm{0}$ using Newton iteration with initial guess $(K_{s_1}+ \delta s \dot{K}_{s_1},\{{\cb\bm{U}}_{s_1}(i) + \delta s \dot{{\cb\bm{U}}}_{s_1}(i)\},\alpha_{s_1} + \delta s \dot{\alpha}_{s_1})$;
      
      \For{$n > 1$}
      \State given $(K_{s_{n}},\{{\cb\bm{U}}_{s_{n}}(i)\},\alpha_{s_n})$ and $(K_{s_{n-1}},\{{{\cb\bm{U}}}_{s_{n-1}}(i)\},{\alpha}_{s_{n-1}})$;
      \State compute an approximate $(\dot{K}_{s_n},\{\dot{{\cb\bm{U}}}_{s_n}(i)\},\dot{\alpha}_{s_n})$ using \eqref{dot_approx2};
      
      \State \begin{varwidth}[t]{0.88\linewidth}\sloppy compute a new equilibrium configuration $(K_{s_{n+1}},\{{\cb\bm{U}}_{s_{n+1}}(i)\},\alpha_{s_{n+1}})$ by solving ${(\bm{F_1},f_K) = \bm{0}}$ using Newton iteration with initial guess ${(K_{s_n}+ \delta s \dot{K}_{s_n},\{{\cb\bm{U}}_{s_n}(i) + \delta s \dot{{\cb\bm{U}}}_{s_n}(i)\},\alpha_{s_n} + \delta s \dot{\alpha}_{s_n})}$. \end{varwidth}
  \EndFor
\end{algorithmic}
\end{algorithm}

\section{Results}
In this section we discuss numerical tests based around applying the pseudo-arclength continuation to both the static and flexible boundary schemes. 

We begin by directly comparing the static boundary scheme and the flexible boundary scheme when applied to a simple toy model, highlighting the superiority of the latter. This is then followed by a study of 
fracture on the $(111)$ cleavage plane in silicon with two interatomic potentials.

\subsection{Mode III toy model}\label{sec:num_toy}
We first consider a toy model of anti-plane Mode III fracture posed on a triangular lattice with lattice constant equal to unity and atoms interacting according to a nearest neighbour pair potential. The total energy is thus of the form
\begin{equation*}
E = \sum_{\substack{i\neq j\\|\bm{\hat{x}}(i)-\bm{\hat{x}}(j)|=1}} \phi(r_{ij}),\;\text{ where } r_{ij} = |x_3(i)-x_3(j)|,
\end{equation*}
where 
\[
\phi(r) = \frac{1}{6}\left(1-\exp(-3r^2)\right).
\]
The resulting material properties are reported in Table~\ref{tab:toy}, including the shear modulus, the surface energy and the Griffith prediction for the critical stress intensity factor $K_G$.

\begin{table}[]
    \centering
    \begin{tabular}{l|c}
     Quantity & Value \\
         \hline
     $a$ & 1.0 \\
     $\mu$ & 3.464 \\
     $\gamma$ & 0.333 \\
     $K_G$ & 0.49501 \\
    \end{tabular}
    \caption{Values of the lattice constant $a$, shear modulus $\mu$, surface energy $\gamma$ and Griffith stress intensity factor $K_G$ computed for the Mode III toy model.}
    \label{tab:toy}
\end{table}

To investigate domain size effects, we consider computational domains of different sizes, each geometrically represented by ball of radius $R$ around the origin {\cb (we refer to Figures \ref{fig:regions} and \ref{fig:regions_ff} for visual insight)}. The three choice of radii are (1) $R=32$, (2) $R=64$ and (3) $R=128$. The fully atomistic  Region {\cb 1} is chosen to consists of all atoms with
\begin{equation}\label{regionI}
|\bm{\hat{x}}(i)| < R - R_{\rm out} - R_{\phi},
\end{equation}
where $R_{\rm out} = 2.1$ corresponds to the width of the annulus of atoms in the far field (Region {\cb 3}), and $R_{\phi} = 1.1$ corresponds to the interaction radius,  specifying the width of the annulus of atoms in the interfacial Region {\cb 2}. As a result in each scheme (1)~$N_1 = 3003$, (2)~$N_1 = 13402$ and (3)~$N_1 = 56500$, respectively.
\subsubsection{Pseudo-arclength continuation with static boundary scheme}
\begin{figure}[!htb]
\includegraphics[width=\columnwidth]{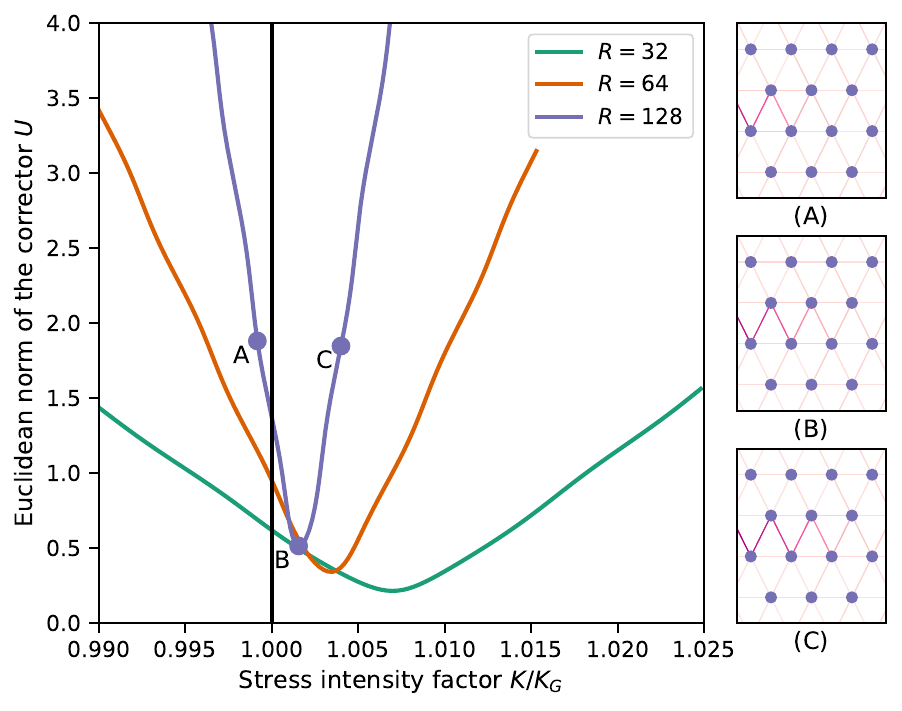}
\caption{\label{fig:static_plot} Solution paths obtained for Mode III toy model using Algorithm \ref{alg1} for three choices of domain size {\corr (the more vertical the path, the greater the domain size)}{\cb , with examples A, B and C of the computed equilibrium configurations, clearly showing that the crack propagates as the solution curve is traversed -- note that this is Mode III in the anti-plane approximation, so the greater the strain on a bond, the {\corr more visible the bond.}}}
\end{figure}

Algorithm \ref{alg1} is first employed to compute solution paths presented in Figure \ref{fig:static_plot}. With no knowledge of the actual crack tip position, the $y$-axis was chosen to represent the Euclidean norm of $\{{\cb\bm{U}}_s(i)\}$.

The plot confirms the intuitively clear notion that $|{\cb\bm{U}}_s|$ will be smallest when there is no mismatch between the predicted crack tip position (in the static boundary scheme fixed at $\alpha = 0$) and the actual crack tip position. Periodic wiggles further indicate a repeating bond-breaking behaviour.

The solution paths are heavily tilted, with no clear range of stress intensity factors for which equilibria exist, as $K$ grows to effectively compensate for $\alpha$ being fixed. In particular, no unstable equilibria are found and the energy is monotonically increasing in $K$, implying that no study of energy barriers is possible. 

An ad-hoc post-processing way of estimating actual values of $\alpha$ and $K$ is to find 
\begin{equation}\label{static-post-process}
\min_{\alpha,K} \left(\sum_{i=1}^{N_*} |K_s\,{\cb\bm{U}}_{\rm CLE}(i) + {\cb\bm{U}}_s(i) - K {\cb\bm{U}}_{\rm CLE}^{\alpha}(i)|^2\right),
\end{equation}
where, in order to avoid boundary effects,  $\{i\}_{i=1}^{N_*}$ corresponds to all atoms such that ${|\bm{\hat{x}}(i)| < \frac{3}{4}R}$. The resulting plots of $\alpha$ against $K$ are shown with dashed lines in Figure \ref{fig:K_alpha_comp}.
\subsubsection{Pseudo-arclength continuation with flexible boundary scheme}\label{sec:flex_standard}
Algorithm \ref{alg2} is now employed to compute solution paths of the toy model for three different domain sizes, as described in Section \ref{sec:num_toy}. 

\begin{figure}[!htb]
\includegraphics[width=1.0\columnwidth]{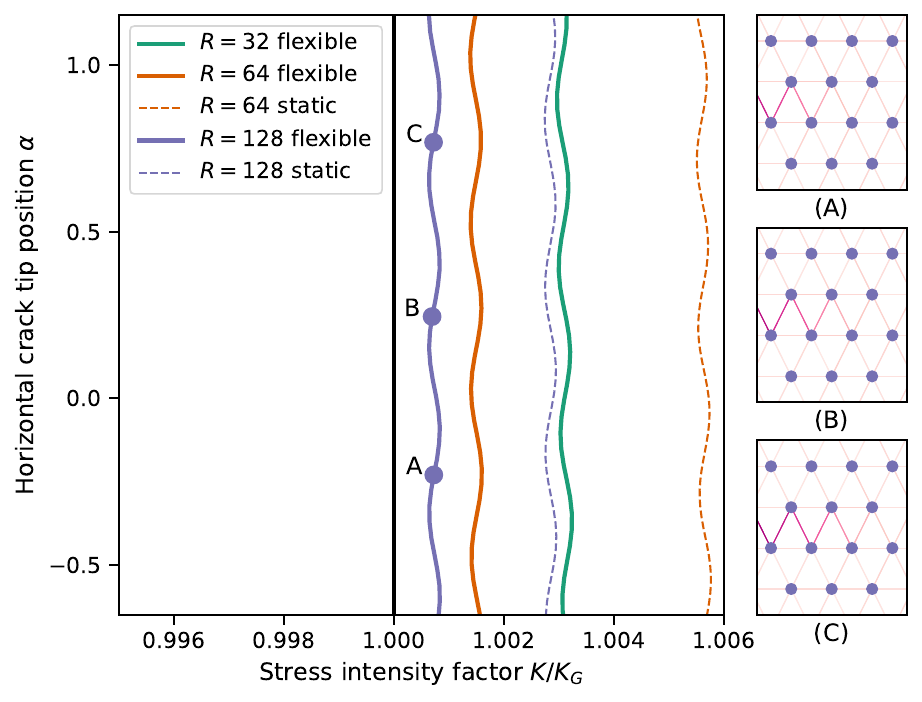}
\caption{\label{fig:K_alpha_comp} Comparison of solutions paths obtained for Mode III toy model using Algorithm \ref{alg2} (solid lines) and Algorithm \ref{alg1} (dotted lines, post-processed via \eqref{static-post-process}) for three domain sizes ({\corr progressing to the left as the domain size grows, with }static $R=32$ case too far away to the right to include). {\cb The insets present examples of the computed configurations, clearly showing crack propagation, with the line intensity convention as in Figure \ref{fig:static_plot}.}}
\end{figure}

A direct comparison of both scheme is shown in Figure \ref{fig:K_alpha_comp}, revealing that the flexible scheme is superior to the post-processing of the static scheme in terms of predicting the range of the stress intensity factors for which equilibria exist. In particular, the flexible scheme employed on a core region with radius $R=32$ is as accurate as the post-processed static scheme employed on a core region with radius $R=128$. 

{\cb Note that in the flexible boundary scheme, outside the tiny interval for $K$ presented in Figure \ref{fig:K_alpha_comp}, no equilibrium configurations exist. Physically this corresponds to loads being large enough for the crack to propagate through the whole material. This is made possible by the adjustment in $\alpha$ --  if $K$ is `strong' enough for the crack to propagate by one lattice spacing from $\alpha$ to $\alpha + 1$ (where the lattice spacing is normalised to $1$), then it will also be strong enough to propagate to $\alpha + 2$ and so on. This of course terminates near the actual boundary of the computational domain, but there the boundary effects are too strong for the model to be in any way meaningful. The finiteness of the computational domain also comes into play in the form a slight tilt of the solution path. }

\begin{figure}[!htb]
\includegraphics[width=1.0\columnwidth]{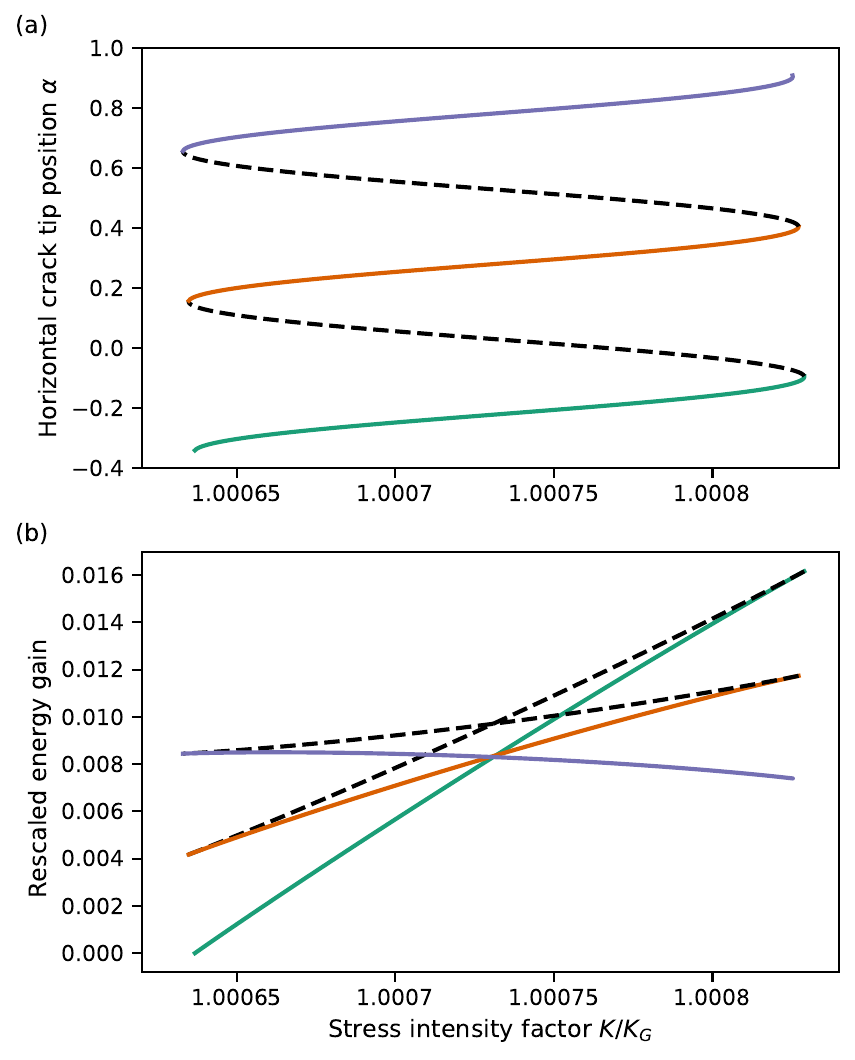}
\caption{\label{fig:flex_barriers} A study of energy barriers in the Mode III toy model for computational domain of radius $R=128${\corr , showing (a) the solution path and (b) the resulting changes in energy.} The rescaled energy gain is $(E - E_*)/E_*$, where $E_*$ is the energy of the bottom left configuration. The dashed parts of the solution path denote index-1 saddle points, which correspond to energetic cost of crack propagation at a given value of $K$, which can be seen by observing in the lower plot that dashed lines lie above their neighbouring solid lines. A nudged elastic band calculation further confirming this being the case is presented in Figure \ref{fig:flex_neb}. The point where stable parts of the solution paths cross corresponds to the critical stress intensity factor $K_c$, notably not quite matching the Griffith stress intensity factor $K_G$. This phenomenon is elaborated upon in Section \ref{sec:error}.}
\end{figure}

With unstable equilibrium configurations corresponding to index-1 saddle points captured in the flexible boundary scheme, a study of energy barriers is now feasible, as showcased in Figure \ref{fig:flex_barriers} and later in Figure \ref{fig:flex_neb}. 

\subsubsection{Predicting the admissible range for $K$}\label{num_pred_k}
\begin{figure}[!htb]
\includegraphics[width=1.0\columnwidth]{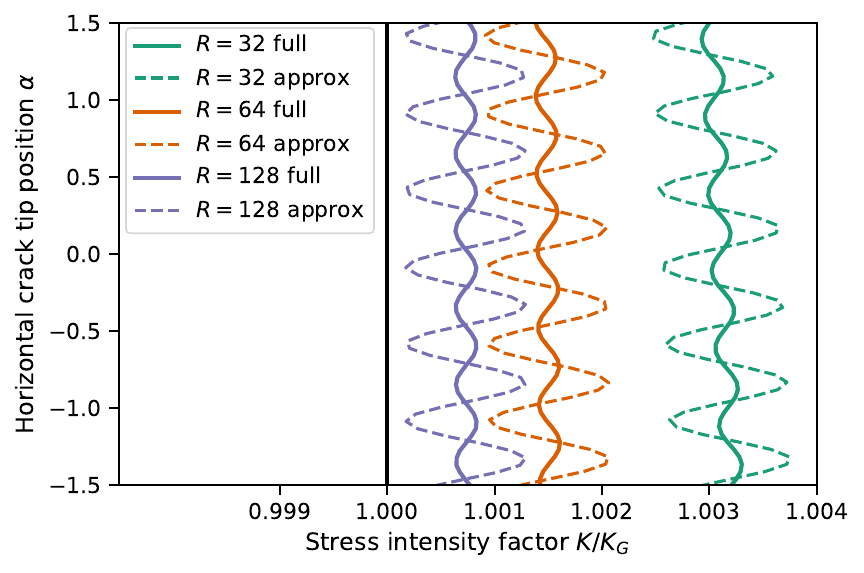}
\caption{\label{fig:K_pred} Solutions paths computed for the Mode III toy model with Algorithm \ref{alg2} (solid lines) for the three domain sizes {\corr (progressing to the left as the domain size grows)}, plotted against a corresponding path of approximate solutions obtained by solving $f_{\alpha}^0 = 0$ from \eqref{f0_alpha} (dashed lines).}
\end{figure}

The ideas developed in Section \ref{sec:pred_K_CLE} are now checked numerically for the toy model presented in Section \ref{sec:num_toy}, again employing three domain sizes. The results are presented in Figure \ref{fig:K_pred}.

The prediction of the range of admissible values of the stress intensity factor based on the CLE displacements only is shown to be fairly accurate, with the magnitude for $K$ matching, while the predicted length of the interval considerably larger than in reality. Importantly, the prediction correctly shifts with the changing domain size, indicating that the range of admissible values for $K$ is to a considerable extent determined by the far-field behaviour only, thus strongly motivating the new formulation of the flexible scheme presented in Section~\ref{sec:new_flex}, which will be tested numerically in the next section.

\subsubsection{Pseudo-arclength continuation with flexible boundary scheme with extended far-field region}\label{sec:num_new_flex}
To test the effect of extending the far-field region discussed in Section~\ref{sec:new_flex}, we consider a computational domain in the form of a ball of radius $\overline{R}=128$ with varying sizes of Region~{\cb 1}. {\cb We refer the reader to the schematic plots in Figures \ref{fig:regions} and \ref{fig:regions_ff} for visual insight.}

As before, the core region is chosen to consists of all atoms satisfying \eqref{regionI}, this time with (1)~$R=8$, (2)~$R=16$, (3)~$R=32$ and (4)~$R=64$. Region~{\cb 2} is again an annulus of width $R_{\phi} = 1.1$ around Region~{\cb 1}. Highlighting the key conceptual change, the width of the outer annulus corresponding to Region~{\cb 3} is now $\overline{R} - R + R_{\rm out}$, as opposed to just $R_{\rm out} = 2.1$ in the standard formulation. As a result $N_3 = 58407$ and in each scheme (1)~$N_1 = 292$ (2)~$N_1 = 1046$, (3)~$N_1 = 3946$ and (4)~$N_1 = 15323$.

A suitably adjusted Algorithm \ref{alg2} is now employed to compute solutions paths. The resulting plots of $K$ against $\alpha$ are presented in the middle panel of Figure \ref{fig:flex1_flex4_error}, which also include the solution path computed with the standard flexible scheme with $R=256$ for comparison. 

The extension of the far-field region drastically increases the accuracy of the flexible boundary scheme, with a tiny fully atomistic region required to have a very accurate prediction for the admissible range of values for the stress intensity factor. This is demonstrated quantitatively in the error analysis in Section \ref{sec:error}.

Despite the large far-field region, the system of nonlinear equations associated with the new scheme when $R=8$ consists of merely $294$ equations, as compared to the standard scheme when $R=128$, which consists of $56502$ equations, thus rendering the new scheme vastly superior. 

\begin{figure}[!htb]
\includegraphics[width=1.0\columnwidth]{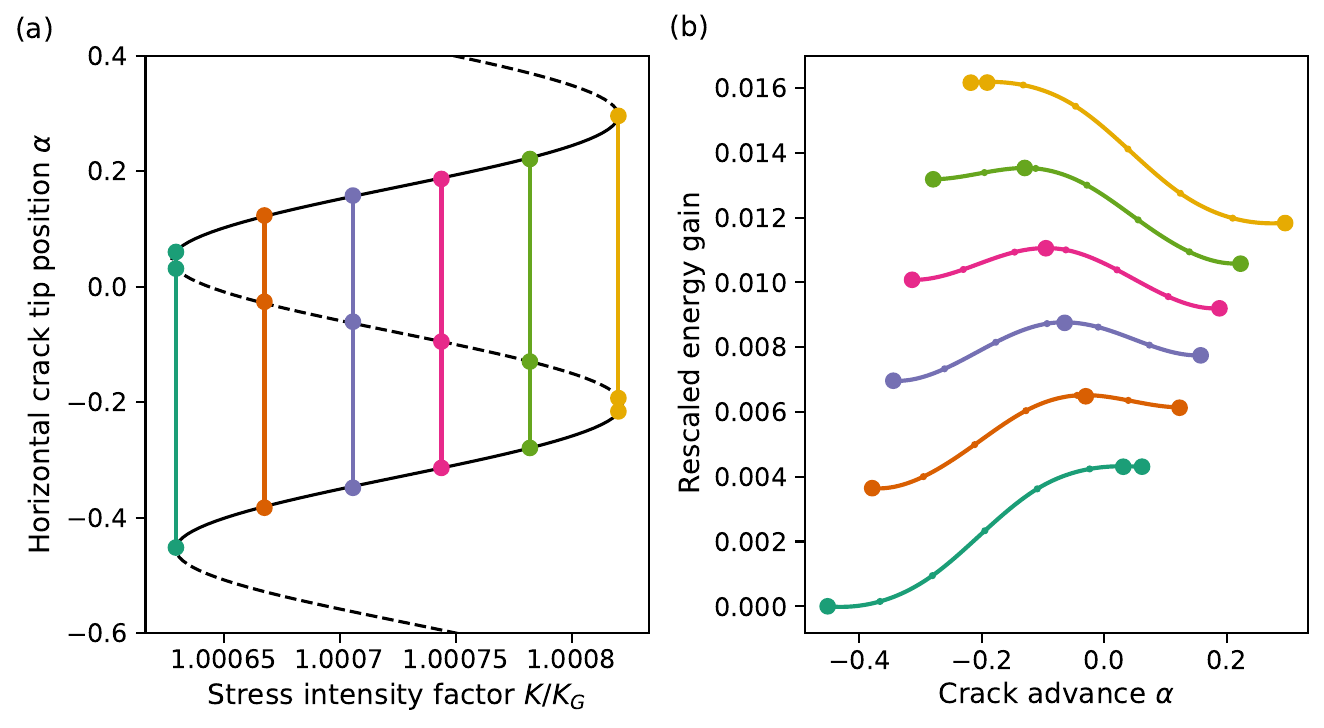}
\caption{\label{fig:flex_neb} A study of energy barriers in the flexible boundary scheme with extended far-field with $R=32$ and $\overline{R} = 128$. {\corr In (a) } the solution path is in black {\corr alternating between a solid and a dashed line}. Six values of $K$ are chosen and in each case an initial minimum energy path (MEP) is formed by linear interpolation between the first stable equilibrium, the saddle in between and the second stable equilibrium. The path is then optimised using the nudged elastic band method~\cite{Makri2019}. The resulting MEPs are shown in {\corr (b) with the leftmost path from (a) corresponding to the lowest energy pathway in (b)}, together with larger dots representing the equilibria computed with pseudo-arclength continuation, thus confirming that the middle equilibrium is indeed a saddle and also confirming lack of other critical points along a given path.}
\end{figure}

Finally, to further confirm that the unstable solutions computed are indeed saddles and that no other critical points can be found along the way, a modified version of the nudged elastic band method~\cite{Makri2019} has been employed on the domain with $R=32$, with details presented in Figure \ref{fig:flex_neb}. 

\subsubsection{Error analysis}\label{sec:error}
\begin{figure*}[!htb]
\includegraphics[width=1.0\linewidth]{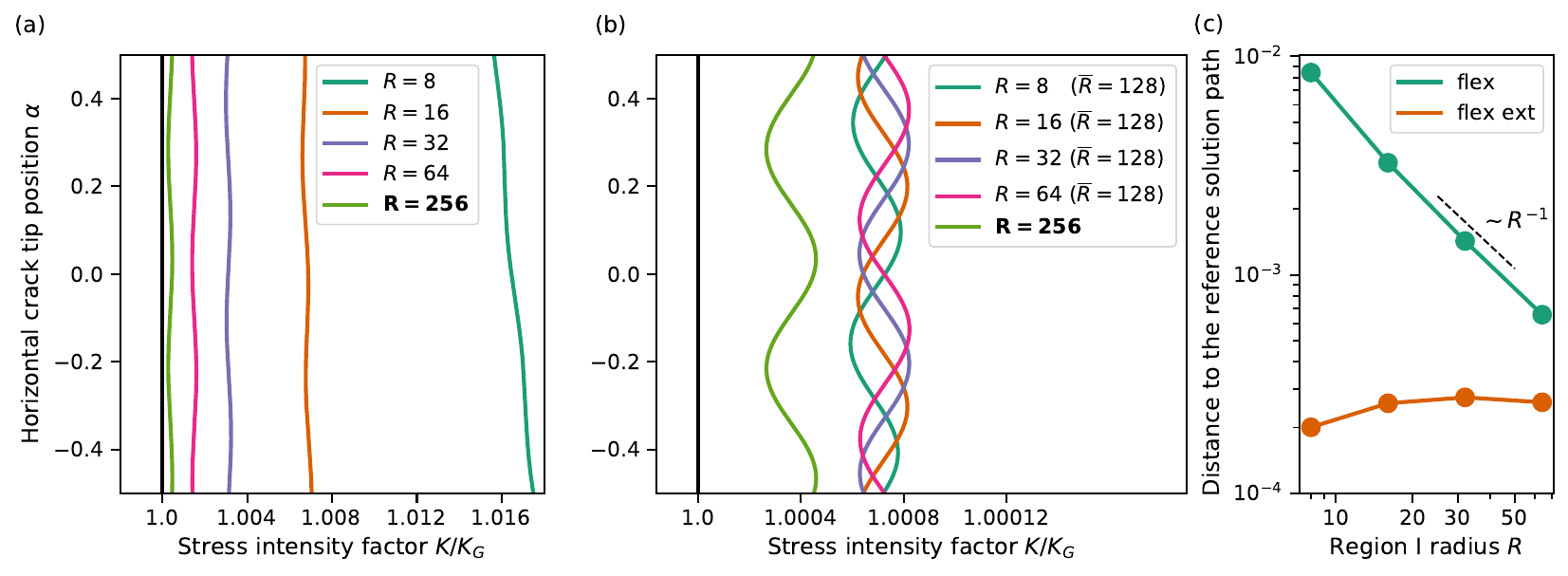}
\centering\caption{\label{fig:flex1_flex4_error} Solution paths and error analysis for (a) flexible and (b) extended far field flexible boundary schemes {\corr (note the change of scale on the $x$-axis between (a) and (b)). In (a) paths move to the left as $R$ increases, but in (b) they remain nearly exactly confined to a fixed interval in $K$}. The {\corr leftmost} solution path {\corr in both (a) and (b) is computed with the standard flexible boundary scheme with $R=256$ and }serves as a reference in the error analysis {\corr shown in (c). The $R^{-1}$ convergence rate is observed for the flexible scheme from (a).} 
}
\end{figure*}

To conclude the numerical investigation of the toy model, a brief error analysis is presented in Figure \ref{fig:flex1_flex4_error}. 
The reference solution path, imitating the infinite limit $N_1 \to \infty$ is obtained with the standard flexible boundary scheme, as described in Section \ref{sec:num_toy}, with $R=256$. Subsequently solution paths obtained with the standard flexible scheme with $R=8,16,32,64$ are computed, as well as solution paths obtained with the extended flexible boundary scheme, as discussed in Section \ref{sec:num_new_flex}, with $\overline{R} = 128$ and $R = 8, 16, 32, 64$. 

The right-hand side plot in Figure \ref{fig:flex1_flex4_error} is produced by computing the Hausdorff distance~\cite{RW98} (intuitively the greatest of all the distances from a point on one line to the closest point on the other line) between a solution path of a given radius and the reference solution path.

Two things are apparent: firstly, the standard flexible scheme yields a rate of convergence of order $O(R^{-1})$, which improves upon a known rate of convergence $O(R^{-1/2})$ of the static scheme proven in Ref.~\onlinecite{2019-antiplanecrack}. A mathematically rigorous proof of the improved rate of convergence will be a subject of further study. Notably, the error analysis together with the study of energy barriers presented in Figure \ref{fig:flex_barriers} and \ref{fig:flex_neb} clearly show that the Griffith prediction for the critical stress intensity factor $K_G$ is only valid in the limit $N_1 \to \infty$.

Secondly, the extended far field flexible boundary scheme remains as accurate as the outer radius, which in the current study is fixed at $\overline{R}=128$. The difference in accuracy is thus most apparent for small values of $R$, confirming the intuition behind this reformulation of the flexible boundary scheme. The underlying reasons for this are also to be explored in a future work. 

\subsection{Mode I fracture of silicon on the $(111)$ cleavage plane}\label{sec:num_real}

We next test our new algorithms on a more complex problem: fracture of silicon on the $(111)$ cleavage plane in the $[11\bar{2}]$ propagation direction. This is known to be the preferred low-energy cleavage orientation, but the precise details of the lattice trapping barriers to brittle fracture remain elusive for the reasons outlined in the introduction, making this a problem of scientific interest as well as an interesting test case.

We consider two interatomic potentials known to give a qualitatively correct description of brittle fracture for this system: modifications of the Tersoff~\cite{Tersoff1988} and Kumagai~\cite{Kumagai2007} potentials, with the interaction length increased and additional screening functions introduced to improve the description of bond-breaking processes~\cite{Pastewka2013}. Without these modifications, neither potential predicts brittle behaviour. The modified potentials have been shown to predict lattice trapping ranges $K_- < K < K_+$ for the $(110)$ cleavage plane in reasonable agreement with DFT, albeit restricted to a small model system with static boundary conditions~\cite{Pastewka2013}, and we thus use them here as a proxy for a fully description of interatomic bonding in silicon. The potentials have not previously been applied to study fracture on the $(111)$ plane, in part because of the complexities introduced by surface reconstructions such as the Pandey $2\times1$ $\pi$-bonded chain~\cite{Kermode2008,Fernandez-Torre2010}, which we do not study here.

A number of small modifications to the FBC method described above are needed. Since analytical Hessians are not readily available for these potentials, we use the finite-difference reformulation of the scheme outlined in Algorithm \ref{alg3}. For comparison with the static case, we also consider a Hessian-free version of Algorithm \ref{alg1}, which can be obtained from Algorithm \ref{alg3} by fixing $\alpha=0$ and $\dot{\alpha}=0$ throughout. The linear elastic predictor ${\cb\bm{U}}^\alpha_\mathrm{CLE}$ and its derivative ${\cb\bm{V}}_\alpha$ are also redefined to account for the anisotropy of the silicon crystal using the near field solution for a crack in a rectilinear anisotropic elastic medium (noting that ${\cb\bm{V}}_\alpha$ can conveniently be obtained from the $xx$ and $xy$ elements of the deformation tensor)~\cite{Sih1965}. The CLE solutions are expanded from two to three dimensions using plane strain loading conditions appropriate for a simulation cell periodic along the crack front line, i.e. $Y_3 = 0$, with the atomistic corrector ${\cb\bm{U}}(i)$ for each atom also becoming three dimensional. In place of the Newton iteration, we solve $({\bm F}_1, f_K) = 0$ with a Newton-Krylov solver as implemented in the \texttt{LGMRES} package~\cite{Baker2005}. 

For large systems, it is necessary to precondition the solver. We used a general purpose preconditioner for materials systems~\cite{Packwood2016}, augmented by a diagonal rescaling of the $f_\alpha$ and $f_K$ components of the preconditioner to balance their magnitudes with that of the atomic forces ${\bm f}(i)$ (as suggested by Sinclair~\cite{Sinclair1975}). Finally, the crack tip force $f_\alpha$ is now computed by summing only over atoms in Region {\cb 2} (or Regions {\cb 2} and {\cb 3} for the extended far-field variant); as discussed after \eqref{falpha} this does not affect the equilibria obtained.
A software implementation of the algorithm is available with the framework of the Atomic Simulation Environment (ASE)~\cite{Larsen2017} as part of the open source \texttt{matscipy} package~\cite{matscipy}. 

To setup the simulations, the lattice and elastic constants and the surface energy of the $(111)$ plane are computed for each potential and are reported in Table~\ref{tab:pots}, along with the Griffith prediction for the critical stress intensity factor $K_G$, obtained using the relaxed surface energy $\gamma_{(111)}$.

\begin{table}[]
    \centering
    \begin{tabular}{l|c|c}
         Quantity & Tersoff+S & Kumagai+S \\
         \hline
     {\corr $r_c$~[\AA{}]}  & 6.0 & 6.0 \\
     {\corr $a$~[\AA{}]} & 5.432 & 5.429 \\
     {\corr $C_{11}$~[GPa]} & 143 & 165 \\
     {\corr $C_{12}$~[GPa]} & 75 & 65 \\
     {\corr $C_{44}$~[GPa]} & 69 & 77 \\
     {\corr $\gamma_{(111)}$~[Jm$^{-2}$]} & 1.20 & 0.89 \\
     {\corr $K_G$~[MPa$\sqrt{\text{m}}$]} & 1.07 & 0.97 \\
    \end{tabular}
    \caption{Values of the cutoff radius $r_c$, lattice constant $a$, cubic elastic constants
    $C_{11}$, $C_{12}$, $C_{44}$, surface energy $\gamma_{(111)}$ and Griffith stress intensity factor $K_G$ computed with the screened versions of the Tersoff and Kumagai interatomic potentials.}
    \label{tab:pots}
\end{table}

Similar to the toy model above, we consider three domain radii (1)~$R=32$~\AA{}, (2)~$R=64$~\AA{} and (3)~$R=128$~\AA{}, with the radius of the fully atomistic Region {\cb 1} chosen to consider atoms with crystal positions
\[
|\bm{\hat{x}}(i)| < R - R_\mathrm{out} - R_\phi
\]
where now we take $R_\mathrm{out} = 2r_c = 12$~\AA{} as the width of the annulus of atoms defining Region {\cb 3} and $R_\phi = r_c = 6.0$~\AA{} for the width of annulus of atoms in the interfacial Region {\cb 2}. For the extended far-field scheme, a further outer annular region of width $r_c$ is added to ensure the forces on atoms in Region {\cb 3} are unaffected by the presence of the outer surface. The corresponding numbers of atoms in Region {\cb 1} are (1) $N_1 = 119$, (2) $N_1 = 1273$, (3) $N_1 = 7286$, respectively. Since we now work in 3D, there are $3N_1+1$ degrees of freedom for the Newton-Krylov solver for the static arc-length calculation, and $3N_1+2$ for the flexible case.

\subsubsection{Pseudo-arclength continuation with the Static and Flexible Boundary Conditions}

\begin{figure}
    \centering
    \includegraphics[width=\columnwidth]{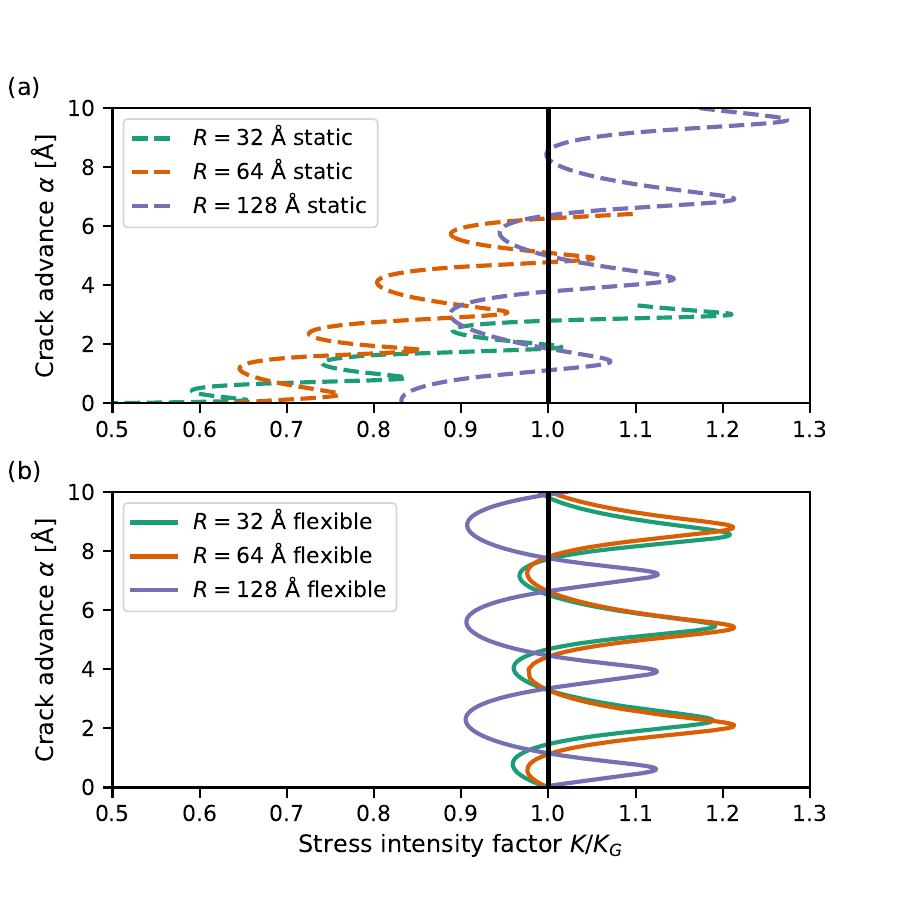}
    \caption{Comparison of solution paths obtained with Algorithm \ref{alg3} for a Si$(111)[11\bar{2}]$ crack modelled with the screened Kumagai potential using {\corr (a) static (dashed lines) and (b) flexible (solid lines) boundary conditions}, for three choices of domain size. For the static cases $\alpha$ and $K$ are obtained by a post-processing fit to the CLE solution.}
    \label{fig:static_vs_flexible}
\end{figure}

\begin{figure*}
    \centering
    \includegraphics[width=\textwidth]{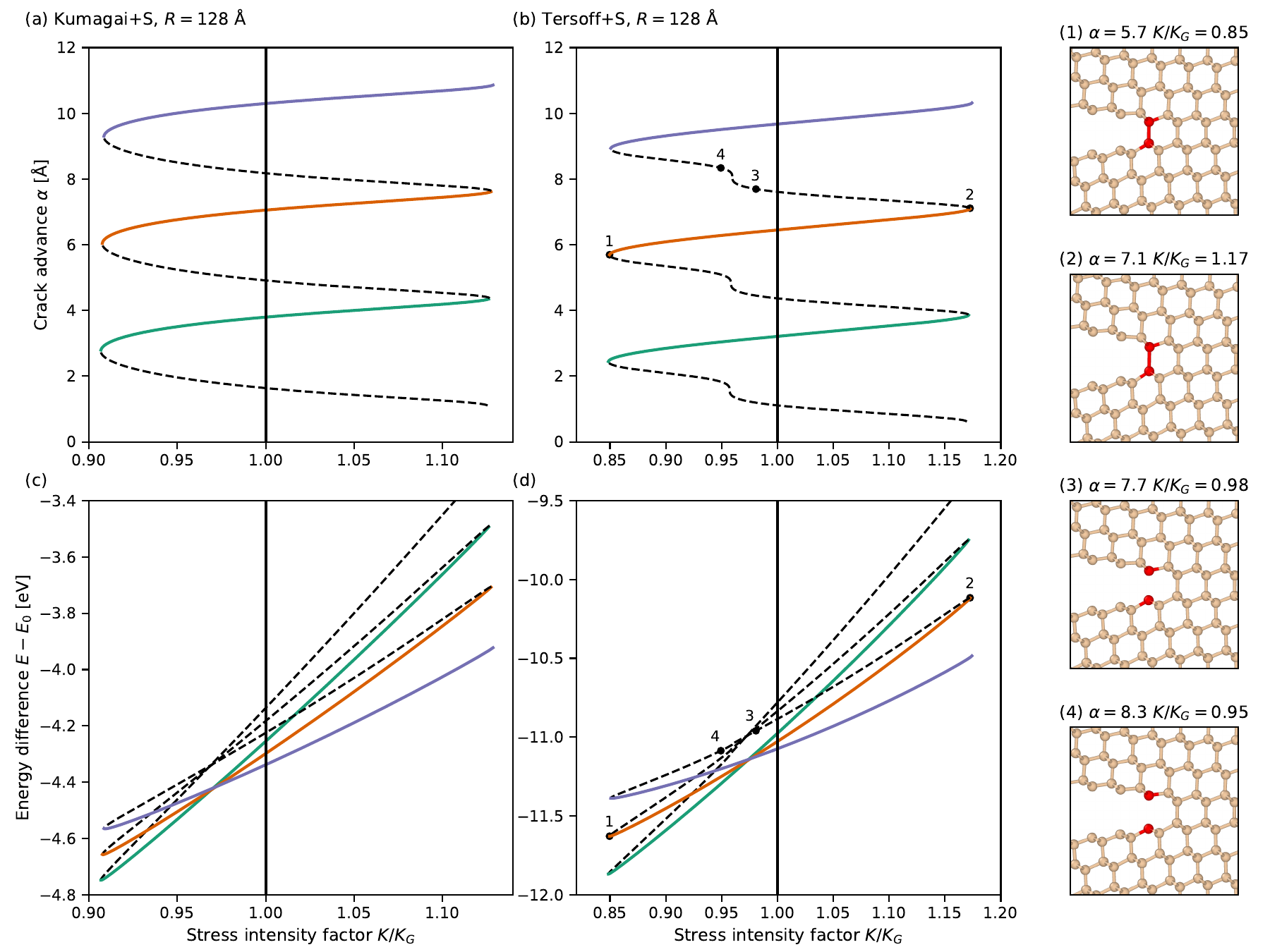}
    \caption{Comparison of lattice trapping of Si$(111)[11\bar{2}]$ cracks predicted by the screened Kumagai {\corr (a, c)} and Tersoff {\corr (b, d)} potentials. Panels {\corr (a) and (b)}: solution paths obtained with Algorithm \ref{alg3}, including stable parts (solid lines) corresponding to energy minima and unstable parts (dashed lines) corresponding to saddle points. Panels {\corr (c) and (d)}: energy difference with respect to the CLE solution with $\alpha=0, K=K_G$. Insets: near-tip atomic positions corresponding to marked points {\corr 1, 2, 3, 4} on the Tersoff solution and energy paths, with the opening bond highlighted in red. }
    \label{fig:kumagai_tersoff}
\end{figure*}

We first perform arc-length continuation calculations with the Kumagai potential for three choices of domain radii, using both static and flexible boundary conditions. The results are shown in Figure~\ref{fig:static_vs_flexible}. 
For the static cases, we employ the post-processing fit for $K$ and $\alpha$ given in \eqref{static-post-process}, leading to the results shown with dashed lines in the figure.
For small domain sizes, the static solutions are highly tilted, while the flexible solutions show the correct periodic behaviour even at the smallest domain size. 
{\cb We note, however, that the finite domain effects are not yet fully understood for the realistic model, such as the apparent significant change from $R=64$ to $R=128$ seen in Figure \ref{fig:static_vs_flexible}. From the purely mathematical point of view, as noted in the concluding section of \cite{2018-antiplanecrack}, it appears that it is not enough to just prescribe $K\hat{\bm{U}}^{\alpha}_{\rm CLE}$ as the boundary condition at infinity --  it should be supplemented by the next order term which behaves like $\sim r^{-1}$ in the strain (this term is absent in the toy model due to the inherent symmetry). {\corr A separate consideration is also needed for near-crack-surface terms, as the continuum models do not account for atomistic surface phenomena -- unless they can be shown to be negligible in comparison with bulk terms, an extra surface far-field term is needed too.} Expanding further upon this in the current work would obscure other useful aspects of the developed method, hence we defer this to future work on this topic.}

{\cb Finite domain effects aside, t}he high accuracy of the flexible scheme allows a careful comparison of the lattice trapping predicted by different choices of interatomic potential to be made, as shown in Figure~\ref{fig:kumagai_tersoff}.
At a domain size of 128~\AA{} the solution paths are already very close to periodic in the crack propagation direction. The energy differences computed with \eqref{energy_decomposition} {\cb (using the far-field approximation from \eqref{E-ff-approx})} illustrated in the lower panels confirm that there is a critical stress intensity factor $K_c$ for which the total energy of the atomistic plus continuum system is equal at all stable energy minima, i.e. before and after crack advance. While the range of lattice trapping $K_- < K < K_+$ predicted by the two potentials is similar, for both potentials $K_c$ is less than the Griffith equilibrium value $K_G$. The values of $K_G$ used here were computed from the relaxed $(111)$ surface energy, indicating, as well as remaining finite size effects, some of the difference could be attributed to local modifications of the surface energy close the crack tip --- a discrepancy that could be further exacerbated by the presence of more complex surface features such as the Pandey $2\times1$ reconstruction. 

The unstable part of the screened Tersoff solution path contains an interesting additional feature around $K = 0.95K_G$. The inset schematics illustrate how this feature arises: moving along the stable path from (A) to (B), the bond at the crack tip remains intact as the centre of the continuum field $\alpha$ advances. The bond gradually opens as we move towards point (C) in the unstable part of the solution path, while between (C) and (D) it opens more rapidly as the atoms `snap' apart. We postulate that this sharp feature is associated with the finite cutoff of the potential, which, despite the screening terms that make fracture simulations feasible, is still a modelling assumption. In future work we aim to compute solution paths with DFT to remove the uncertainty associated with the use of simplified potentials: this remains out of reach for the present since, despite the considerable improvements in accuracy afforded by the flexible scheme, converged solution paths still require a large number of force evaluations on systems comprising several thousand atoms.

\subsubsection{Predicting the admissible range of $K$}
\begin{figure}
    \centering
    \includegraphics[width=\columnwidth]{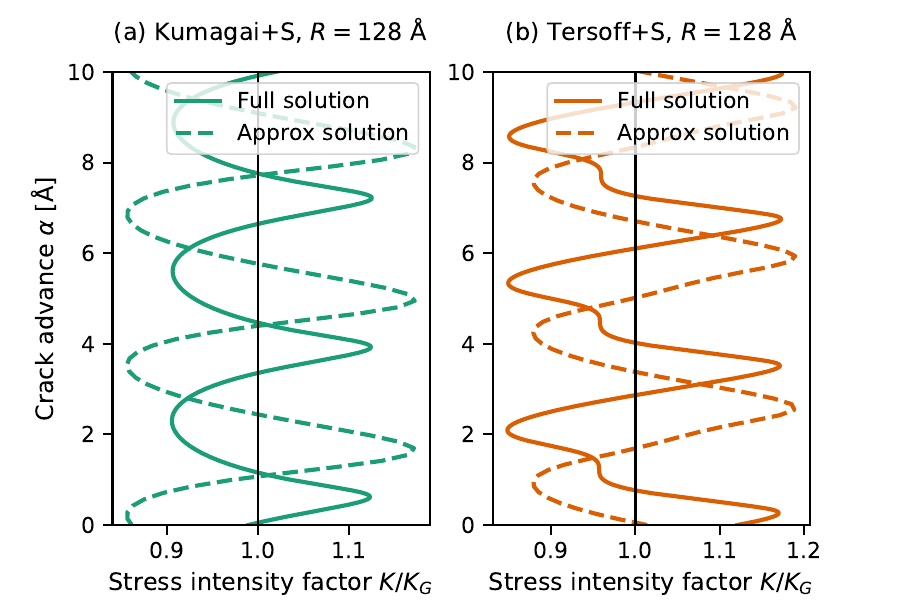}
    \caption{Comparison of full solution paths obtained by arc-length continuation with Algorithm \ref{alg3} (solid lines) and corresponding approximate solution paths for $f^0_\alpha = 0$ (dashed lines) for a Si$(111)[11\bar{2}]$ crack modelled with {\corr(a)  screened Kumagai and (b) screened Tersoff potentials}, using a domain radius of 128~\AA{}.}
    \label{fig:full_vs_approx}
\end{figure}

The admissible range of $K$ is now predicted by {\cb the simple procedure introduced in Section \ref{num_pred_k}, i.e. by} finding roots of equation \eqref{f0_alpha}, namely $f^0_{\alpha}(K, \alpha) = 0$, leading to the predictions shown with the dashed lines in Figure~\ref{fig:full_vs_approx}{\cb , compared against full solution paths computed with pseudo-arclength continuation with flexible boundary.} Here, $K$ is found numerically for each value of $\alpha$ in a 200-element grid.

For both potentials, the admissible range of $K$ is in reasonable agreement with that computed in the full solution paths, suggesting that our approach provides a useful way to estimate the stable range of $K$ for the cost of a fixed number of force evaluations on the full domain.

\subsubsection{Pseudo-arclength continuation with an extended far-field region}
To conclude the numerical tests, we also apply the extended far-field scheme of \eqref{F2} to the Si$(111)[11\bar{2}]$ crack system, modelled using the screened Tersoff potential. 
The overall domain size is fixed at $\bar{R}=128$~\AA{}, and two choices of radii for Region {\cb 1} are considered: $R_I = 14$~\AA{} and $R_I = 46$~\AA{}, chosen since these lead to problems with the same numbers of degrees of freedom as the $R=32$~\AA{} and $R=64$~\AA{} flexible models considered earlier.

\begin{figure}
    \centering
    \includegraphics[width=\columnwidth]{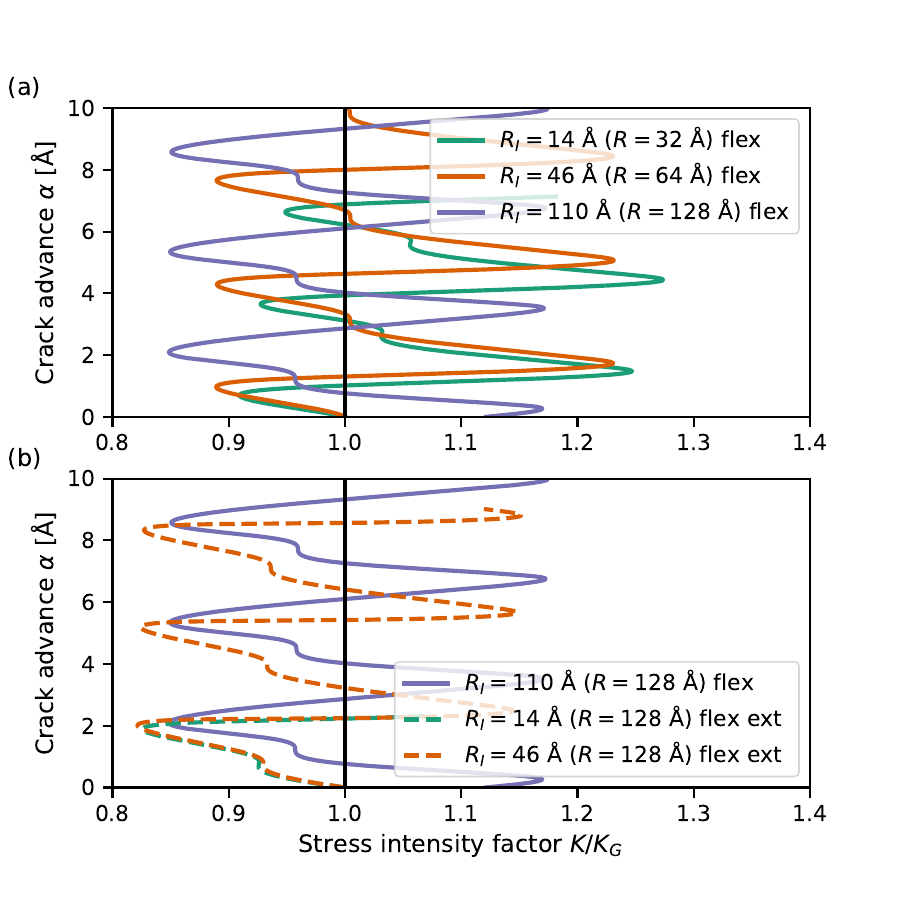}
    \caption{Comparison of original (denoted `flex', {\corr panel (a)}, solid lines) and extended far field (denoted `flex ext', {\corr panel (b)}, dashed lines) variants of the flexible boundary condition approaches to pseudo-arclength continuation in an Si$(111)[11\bar{2}]$ crack system modelled with the screened Tersoff potential. The $R=128$~\AA{} `flex' result is shown in both panels to allow comparison. Calculations with the extended scheme show improved accuracy at smaller domain radii.}
    \label{fig:extended_far_field}
\end{figure}

The results are illustrated in Figure~\ref{fig:extended_far_field}. Although there is an improvement over the standard flexible scheme in convergence towards the reference $R=128$~\AA{} solution path, particularly for the smallest Region {\cb 1} size, these results do not provide convincing evidence that a larger far-field region significantly enhances the accuracy of the scheme. This in in contrast to the results obtained with the toy model, {\cb again} suggesting that an enhanced {\cb far-field} predictor that improves the match with the atomistic model is needed to further increase accuracy. {\cb This will be explored in a separate work.}

\section{Conclusions}

In this work we have reported an extension of Sinclair's flexible boundary condition algorithm to allow full solution paths for cracks to be computed using pseudo-arclength continuation. We have also introduced an extension of the FBC algorithm which allows information to be incorporated from a larger far-field region, and which also provides a stepping stone towards putting the method on a more rigorous mathematical footing. We demonstrated the approach for Mode III fracture with a 2D toy model, and for Mode I fracture of silicon using realistic interatomic potentials that give a qualitatively correct description of fracture. 

In future, our approach will enable a detailed study of lattice trapping barriers to brittle fracture to be carried out using increasingly realistic models of interatomic bonding, going beyond the screened bond-order potentials demonstrated here, for example by using machine-learning interatomic potentials~\cite{Bartok2018} or DFT directly.
This could help to resolve questions such as the role of blunt-sharp-blunt crack tip reconstruction observed during fracture in the Si$(110)[1\bar{1}0]$ crack system~\cite{Swinburne2017}, where NEB calculations demonstrated the crack is bluntened at stable minima and sharp at the unstable transition states.
Moreover, the new approach could be expanded to study crack path selection, known to exhibit complex phenomenon in anisotropic materials~\cite{Mesgarnejad2020}, or the dynamics of three dimensional crack fronts, going beyond previous work that was limited to simple interatomic potentials and small model systems~\cite{Kermode2015}. 
Truly accurate predictions of critical stress intensity factors and lattice trapping ranges require a quantum mechanical approach, at least near the crack tip. Hybrid schemes such as QM/MM (quantum mechanics/molecular mechanics), previously applied to dynamic fracture~\cite{Kermode2008}, could be combined with the algorithms introduced here to make quantitative fracture toughness calculations accurate and affordable. {\corr A potential route to extending the NCFlex scheme to the QM/MM framework is to use a buffered QM/MM approach presented in \cite{2015-qmtb2}, allowing to converge force accuracy with respect to the buffer width.}
Established routes could then be followed to produce atomistically informed continuum models~\cite{Moller2013,Tahir2013,Moller2018}.

Before this can be done, however, further work is needed to assess finite-size effects. For the silicon fracture application, we have demonstrated that the flexible scheme is superior to static boundaries, but not yet quantified the convergence rate, meaning that the new algorithms cannot yet be used for predictive materials science. Ultimately, it is hoped that the flexible boundary scheme and numerical continuation techniques can be combined with higher-order far-field predictions to increase accuracy in a quantifiable manner. 

Finally, we note that the pseudo-arclength continuation used here would also be applicable to other defects such as dislocations by replacing the stress intensity factor $K$ as a bifurcation parameter with the applied shear stress, which also enters as a prefactor in front of the CLE solution.

\acknowledgements

We thank Christoph Ortner and Lars Pastewka for useful discussions.
We acknowledge funding from the EPSRC under grant numbers EP/R012474/1, EP/R043612/1 and EP/S028870/1. Additional support was provided by the Leverhulme Trust under grant RPG-2017-191 and the Royal Society under grant number RG160691.
The authors would like to acknowledge the University of Warwick Scientific Computing Research Technology Platform for assistance in the research described in this paper.


\appendix*
\section{Computation of tangents in the pseudo-arclength continuation scheme}\label{app1}
In the static boundary scheme given by $\bm{F_1} = 0$, differentiating both sides with respect $s$ yields
\begin{equation}\label{dsF0}
\bm{0} = H_{s_n} \dot{{\cb\bm{U}}}_{s_n} + \dot{K}_{s_n} \bm{b}_{s_n}^K,
\end{equation}
where
\begin{equation}\label{Hs}
\left(H_{s_n} \dot{{\cb\bm{U}}}_{s_n}\right)(i) = \sum_{j=1}^{N_1} H_{s_n}(i,j) \cdot \dot{{\cb\bm{U}}}_{s_n}(j).
\end{equation}
Here $H_{s_n}(i,j)$ is $(i,j)$-th entry of the Hessian operator evaluated at $s_n$. In an infinite crystal, this is an infinite block matrix with 
\[
H_{s_n}(i,j) = \frac{\partial^2 E}{\partial \bm{x}_{s_n}(i)\partial \bm{x}_{s_n}(j)},
\]
with a short-hand notation $H_{s_n}$ used to denote its part related to atoms in Region {\cb 1}, which is thus a $N_1 \times N_1$ block matrix.

The other term on the right-hand side of \eqref{dsF0} is given by
\[
\bm{b}^{K}_{s_n}(i) = \sum_{j=1}^{N_2} H_{s_n}(i,j) \cdot {\cb\bm{U}}^{\alpha_{s_n}}_{\rm CLE}(j),
\]
where crucially the summation here is over both the core and the interface regions (thus the Hessian operator here is effectively a rectangular block matrix of size $N_2 \times N_1$), whereas in \eqref{Hs} the summation is only over the core region. 

It follows from \eqref{s-arc} (with $\dot{\alpha} \equiv 0$ in the static boundary scheme) and \eqref{dsF0} that
\begin{equation}\label{udot_kdot}
\dot{{\cb\bm{U}}}_{s_n} = -\dot{K}_{s_n}\left(H_{s_n}^{-1} \bm{b}_{s_n}^K\right),\quad \dot{K}_{s_n} = \pm\left(|H_{s_n}^{-1} \bm{b}_{s_n}^K|^2 + 1\right)^{-1/2},
\end{equation}
provided the square block matrix $H_{s_n}$ is invertible. The case when it is not invertible is known as a bifurcation point and it will be discussed below.

In the flexible boundary scheme given by $\bm{F_1} = 0$ differentiating with respect to $s$ implies 
\begin{equation}\label{dsF1}
    \begin{cases}
    \bm{0} &=\; H_{s_n} \dot{{\cb\bm{U}}}_{s_n} + \dot{K}_{s_n} \bm{b}_{s_n}^K + \dot{\alpha}_{s_n} \bm{b}_{s_n}^{\alpha}\\
    0 &=\; \bm{b}^{\alpha}_{s_n} \cdot \dot{{\cb\bm{U}}}_{s_n} + \dot{K}_{s_n} C_{s_n}^{\alpha,K} + \dot{\alpha}_{s_n} C_{s_n}^{\alpha,\alpha},
    \end{cases}
\end{equation}
where
\begin{align*}
\bm{b}_{s_n}^{\alpha}(i) &= \sum_{j=1}^{N_2}H_{s_n}(i,j)\cdot K\,{\cb\bm{V}}_{\alpha_{s_n}}(j),\\ 
C_{s_n}^{\alpha,K} &= \sum_{i=1}^{N_2}\left( \bm{b}_{s_n}^{K}(i) \cdot K\, {\cb\bm{V}}_{\alpha_{s_n}}(i) + \bm{f}(i) \cdot {\cb\bm{V}}_{\alpha_{s_n}}(i) \right),\\
C_{s_n}^{\alpha,\alpha} &=  \sum_{i=1}^{N_2} \left(\bm{b}_{s_n}^{\alpha}(i) \cdot K\, {\cb\bm{V}}_{\alpha_{s_n}}(i) + \bm{f}(i) \cdot K\, {\cb\bm{V}}^{(2)}_{\alpha_{s_n}}(i)\right),
\end{align*}
with ${\cb\bm{V}}^{(2)}_{\alpha} = -\partial_1 {\cb\bm{V}}_{\alpha}$.

Note that \eqref{dsF1} applies to the newly formulated scheme ${\bm{F_2} = \bm{0}}$ as well, except that the sums defining $\bm{b}_{s_n}^{K}(i)$, $\bm{b}_{s_n}^{\alpha}(i)$, $C_{s_n}^{\alpha,K}$ and $C_{s_n}^{\alpha,\alpha}$ should be over $i=1,\dots,N_3$.

If $H_{s_n}$ is invertible, then equations \eqref{s-arc} and \eqref{dsF1} together imply that 
\begin{align}\label{udot_kdot_alphadot}
\dot{K}_{s_n} = \pm\left(|A_3|^2 + \frac{A_2}{A_1} + 1\right)^{-1/2},\;
\dot{{\cb\bm{U}}}_{s_n} = \dot{K}_{s_n}\,A_3,\; \dot{\alpha}_{s_n} = \dot{K}_{s_n}\,\frac{A_2}{A_1},
\end{align}
where
\begin{align*}
A_1 &= \sum_{i=1}^{N_2}-\left(H_{s_n}^{-1} \bm{b}_{s_n}^{\alpha}(i)\right) \cdot b_{s_n}^{\alpha}(i) + C_{s_n}^{\alpha,\alpha},\\
A_2 &= \sum_{i=1}^{N_2}\left(H_{s_n}^{-1} \bm{b}_{s_n}^{K}(i)\right) \cdot b_{s_n}^{\alpha}(i) - C_{s_n}^{\alpha,K},\\
A_3 &= -\frac{A_2}{A_1}H_{s_n}^{-1} \bm{b}_{s_n}^{\alpha} - H_{s_n}^{-1} \bm{b}_{s_n}^{K}. 
\end{align*}

With $(K_{s_n}, \{{\cb\bm{U}}_{s_n}(i)\},\alpha_{s_n})$ and $(\dot{K}_{s_n}, \{\dot{{\cb\bm{U}}}_{s_n}(i)\},\dot{\alpha}_{s_n})$ known, a standard Newton iteration with initial guess
\[
(K_{s_n}, \{{\cb\bm{U}}_{s_n}(i)\},\alpha_{s_n}) + \delta s(\dot{K}_{s_n}, \{\dot{{\cb\bm{U}}}_{s_n}(i)\},\dot{\alpha}_{s_n}),
\]
is guaranteed to converge to a new solution 
$(K_{s_2}, \{{\cb\bm{U}}_{s_2}(i)\},\alpha_{s_2})$ satisfying $(\bm{F}_i,f_K) = \bm{0}$ provided $\delta s$ is small enough (see Figure \ref{fig:schematic_plot} for visual insight behind this). 

Furthermore, the $s$ derivative $(\dot{K}_{s_2}, \{\dot{{\cb\bm{U}}}_{s_2}(i)\},\dot{\alpha}_{s_2})$ can now be handily computed with an approximate finite-difference-like scheme, given, for $\bm{F_0} = \bm{0}$, by
\begin{subequations}\label{dsF0_approx}
\begin{align}
    \bm{0} &= H_{s_n}\dot{{\cb\bm{U}}}_{s_2} + \dot{K}_{s_2}\bm{b}_{s_n}^K,\\
    1 &= \sum_{i=1}^{N_1}\dot{{\cb\bm{U}}}_{s_n}(i)\cdot \dot{{\cb\bm{U}}}_{s_2}(i) + \dot{K}_{s_n}\dot{K}_{s_2}
\end{align}
\end{subequations}
and, for $\bm{F_1} = \bm{0}$ (and also for $\bm{F_2} = \bm{0}$ after adjusting limits of summation), by
\begin{subequations}\label{dsF1_approx}
\begin{align}
    \bm{0} &= H_{s_n}\dot{{\cb\bm{U}}}_{s_2} + \dot{K}_{s_2}\bm{b}_{s_n}^K + \dot{\alpha}_{s_2} \bm{b}_{s_n}^{\alpha},\\
    0 &= \bm{b}^{\alpha}_{s_n} \cdot \dot{{\cb\bm{U}}}_{s_2} + \dot{K}_{s_2} C_{s_n}^{\alpha,K} + \dot{\alpha}_{s_2} C_{s_n}^{\alpha,\alpha},\\
    1 &= \sum_{i=1}^{N_1}\dot{{\cb\bm{U}}}_{s_n}(i)\cdot \dot{{\cb\bm{U}}}_{s_2}(i) + \dot{K}_{s_n}\dot{K}_{s_2},
\end{align}
\end{subequations}
It is a standard assertion of bifurcation theory~\cite{cliffe_spence_tavener_2000} that the linear systems of equations given by \eqref{dsF0_approx} or \eqref{dsF1_approx}  remain solvable even at the points where stability change, corresponding to cases when $H_{s_n}$ is not invertible, thus allowing us to traverse full the full bifurcation diagram.


\bibliography{refs}
\end{document}